\documentclass[]{emulateapj}
\usepackage{natbib,amsmath}
\shorttitle{The Keck+Magellan Survey for LL Absorption II.}
\shortauthors{Prochter et al.\ }

\begin{document}

\newcommand{\ohr}{O$^0$/H$^0$}
\newcommand{\aslls}{SLLS$_a$}
\newcommand{\bslls}{SLLS$_b$}
\newcommand{\cslls}{SLLS$_c$}
\newcommand{\mnh}{N_{\rm H}}
\newcommand{\mnhi}{N_{\rm HI}}
\newcommand{\nhi}{$N_{\rm HI}$}
\newcommand{\lya}{Ly$\alpha$}
\newcommand{\lyb}{Ly$\beta$}
\newcommand{\lyg}{Ly$\gamma$}
\newcommand{\lyd}{Ly$\delta$}
\def\sci#1{{\; \times \; 10^{#1}}}
\newcommand{\kms}{km~s$^{-1}$ }
\def\N#1{{N({\rm #1})}}
\newcommand{\cm}[1]{\, {\rm cm^{#1}}}
\newcommand{\mkms}{{\rm \; km\;s^{-1}}}
\newcommand{\delv}{\Delta v}
\newcommand{\lofx}{$\ell_{\rm Mg}(X)$}
\newcommand{\lofz}{$\ell_{\rm Mg}(z)$}
\newcommand{\mlofx}{\ell_{\rm Mg}(X)}
\newcommand{\mlofz}{\ell_{\rm Mg}(z)}
\newcommand{\msol}{M_\odot}
\def\fnhi{$f(\mnhi)$}
\def\mfnhi{f(\mnhi)}
\def\qso{PKS2000$-$330}
\def\smm{\sum\limits}
\def\fciv{$f_{\rm CIV}^{\rm cool}$}
\def\mfciv{f_{\rm CIV}^{\rm cool}}

\title{The Keck $+$ Magellan Survey for Lyman Limit Absorption II:  
A Case Study on Metallicity Variations}

\author{Gabriel E. Prochter\altaffilmark{1}, 
J. Xavier Prochaska\altaffilmark{1},
John M. O'Meara\altaffilmark{2,3}, 
Scott Burles\altaffilmark{3},
Rebecca A. Bernstein\altaffilmark{1} 
\\}
\altaffiltext{1}{Department of Astronomy and Astrophysics, UCO/Lick Observatory, 
University of California, 1156 High Street, Santa Cruz, CA 95064}
\altaffiltext{2}{Department of Chemistry and Physics, Saint Michael's College.
One Winooski Park, Colchester, VT 05439}
\altaffiltext{3}{Visiting Astronomer, Las Campanas Observatory}
%\altaffiltext{4}{}

\begin{abstract}
We present an absorption line analysis of the Lyman limit
system (LLS) at $z \approx 3.55$ in our Magellan/MIKE spectrum
of \qso.  Our analysis of the Lyman limit and
full \ion{H}{1} Lyman series constrains the total 
\ion{H}{1} column density of 
the LLS ($\mnhi = 10^{18.0 \pm 0.25} \cm{-2}$ 
for $b_{\rm HI} \ge 20 \mkms$)
and also the \nhi\ values of the velocity subsystems 
comprising the absorber.  We measure ionic column densities for 
metal-line transitions associated with the subsystems and
use these values to constrain the ionization
state ($>90\%$ ionized) and relative abundances of the gas.
We find an order of magnitude dispersion in the metallicities
of the subsystems, marking the first detailed analysis of 
metallicity variations in an optically thick absorber.
The results indicate that metals are not well mixed within
the gas surrounding high~$z$ galaxies.  
Assuming a single-phase photoionization model,
we also derive an $\mnh$-weighted
metallicity, $\rm <[Si/H]> = -1.66 \pm 0.25$, which matches
the mean metallicity in the neutral ISM in high $z$ damped
\lya\ systems (DLAs).  Because the line density of LLSs is over
$10\times$ higher than the DLAs, we propose that the
former dominate the metal mass-density at $z \sim 3$ and that
these metals reside in the galaxy/IGM interface.
Considerations of a multi-phase model do not qualitatively change
these conclusions.
Finally, we comment on an anomalously large O$^0$/Si$^+$ ratio
in the LLS that suggests an ionizing radiation field
dominated by soft UV sources (e.g.\ a starburst galaxy).
Additional abundance analysis is performed on the 
super-LLS systems at $z \approx 3.19$.
\end{abstract}


\section{Introduction}

Lyman limit systems (LLSs) are the 
`clouds' along quasar sightlines that have sufficient
\ion{H}{1} column density \nhi\ to be optically
thick ($\tau > 1$) at the Lyman limit, i.e.\ $\mnhi \gtrsim 10^{17.2} \cm{-2}$.
This definition separates LLSs
from the \lya\ forest lines (clouds with typical $\mnhi < 10^{15} \cm{-2}$)
which trace the intergalactic medium \citep[IGM;][]{rau98} 
and the damped \lya\ systems (DLAs; $\mnhi \ge 2 \sci{20} \cm{-2}$) that trace high
redshift protogalaxies \citep{wgp05}.  On the basis of
their intermediate \ion{H}{1} surface density, one may expect the LLSs to 
represent the physical interface between the IGM and high $z$ galaxies.
%This interface is critical
%to processes of high $z$ galaxy formation because it identifies
%gas densities associated with 
%where gas flows into and out of galaxies.
Studies of the LLSs, therefore, may constrain the accretion
history of galaxies and the transport of material back to the IGM.

%Presently, the LLSs are the least explored set of the three
%\ion{H}{1} classes of high $z$ absorption systems.
%This is partly an observational bias.  Although it is 
%straightforward to measure the incidence of LLSs along
%quasar sightlines \citep[the number per unit redshift is 
%$\ell(z) \approx 2$ at $z = 3$;][]{storrie},
%analysis of the physical conditions in this gas is
%challenged by saturation of the low-order Lyman-series
%lines (e.g.\ \lya, \lyb).
%Furthermore, the gas is generally
%photoionized and one infers many physical properties by
%estimating ionization corrections to observed ionic column
%densities \citep[e.g.][]{p99}.

%A fundamental measure of quasar absorption line systems is the
%chemical enrichment of the gas.  
A key tracer of these processes is the metals intermixed with the gas.  
These are presumably formed in the deep potential wells
of galaxies, imported to the interstellar medium by stellar winds
and supernova explosions, and (possibly) carried to greater
distances by tidal disruption and/or star-formation and 
AGN feedback processes.
With the advent of the 10m Keck~I telescope coupled with the
HIRES echelle spectrometer \citep{vogt94}, observers 
demonstrated that a significant fraction of the IGM is 
enriched in heavy elements of C, Si, and O \citep{tytler95,csk+95}.
More recent studies have established the
variation of metallicity with density and redshift
\citep{schaye03,simcoe04,ads+08} and demonstrate a median metallicity of
roughly 1/1000 solar. These observations constrain the integrated
enrichment by galaxies of their surrounding medium \citep[e.g.][]{ahs+01}.
Studies of the metallicity distribution and evolution in the
damped \lya\ systems \citep{pettini94,pgw+03}, meanwhile, constrain the
star-formation history of high $z$ galaxies and the processes of ISM
enrichment \citep{mirka04,je06}.  
These studies demonstrate that the ISM of most high $z$ galaxies
is metal-poor ($\approx 1/30$ solar), that none have a metallicity
less than 1/1000 solar, and that a wide dispersion (100$\times$)
exists from galaxy to galaxy \citep{pgw+03}.

Although precise measurements for over 100 damped \lya\ galaxies 
have been acquired \citep{pgw+03,ledoux06,pwh+07},  
constraints on metallicity variations
within single galaxies are very limited.
The metal-line transitions generally show absorption by
multiple components spanning $\approx 100 \mkms$ \citep{pw97},
presumably by gas distributed on galactic (kpc) scales.
The Lyman series, however, is in general too saturated to 
yield the \ion{H}{1} column densities of the velocity components
identified in the metal-line transitions.
Therefore, these data provide little constraint on 
metallicity gradients along the sightline
even as a function of velocity much less position.
The studies to date have focused instead on 
relative abundance ratios within DLAs which best trace variations
in the gas density, depletion, and/or ionization state
\citep{pw96,lrd+02,pro03,dz06}.  

In contrast to DLAs, the higher order Lyman series transitions of LLSs are
often unsaturated and one may (in principal) resolve the \ion{H}{1} absorption
into multiple velocity components, each with a well-measured 
\ion{H}{1} column density.  It would then be possible to explore 
variations in the physical conditions of these components,
including the gas metallicity.  This analysis has not been extensively
pursued, primarily because of the significant observational demands:
one requires high S/N, echelle observations covering the full Lyman series.
This demands wavelength coverage $\lambda < 4000$\AA\ for sources at $z<3.4$
which is observationally challenging.
The analysis, as well, is both time-consuming and limited by 
the uncertainties of photoionization modeling. 
With only a few exceptions \citep[e.g.][]{pb99}, the only LLSs
analyzed thus far are the rare and special subset of LLS that
are sufficiently quiescent kinematically to permit a measurement
of the D/H ratio \citep{bt_1937,bkt99}.

Researchers have focused instead on the integrated
metallicities of the so-called super-LLS (SLLS; also referred to
as sub-DLAs) with $\mnhi \ge 10^{19} \cm{-2}$ 
\citep{p99,peroux_slls03,poh+06}.
The \lya\ damping wings for these systems are resolved by high-resolution
spectroscopy and therefore allow a precise constraint on their total
\nhi\ value.  In turn, the gas metallicity may be estimated subject
to photoionization corrections.  It is rare, however, that the full
Lyman series is observed or analyzed, and absent the \nhi\ component
structure one cannot study metallicity gradients.

This paper marks the second in a series presenting results on
a survey of $z \sim 3$ Lyman limit systems.  In our first paper
\citep[][; hereafter Paper~I]{opb+07}, we presented constraints
on the \ion{H}{1} frequency distribution of the SLLSs
absorbers with $\mnhi = 10^{19} \cm{-2}$ to $10^{20.3} \cm{-2}$, from
a survey of echelle and echellette data.  In this paper,
we present our first detailed analysis of a Lyman limit system
with $\mnhi \approx 10^{18} \cm{-2}$.  We demonstrate that 
one can precisely constrain the \ion{H}{1} column densities
of the metal-bearing `clouds' in such systems when provided with 
high-resolution spectroscopy covering the full Lyman series.
Furthermore, we 
measure the ionization state, metallicity, and relative abundances
of the gas through comparisons with models of ionization equilibrium.
In this fashion, we can also explore metallicity variation within the LLS,
presumably corresponding to galactic (sub-Mpc) scales.

Our target is \qso, a bright ($m_v = 17.3$) quasar at
$z_{em} = 3.77$ \citep{psj+82} discovered by \cite{jbg+82}.
Several papers on the 
subject of absorption systems have targeted this quasar 
previously \citep{hunstead86,mhp+86}.
These authors characterized the \lya\ forest 
and the properties of several strong absorption systems along the sightline.  
In their spectra the authors describe four metal-line systems with
\ion{H}{1} absorption indicative of LLSs.  Two are a 
close pair at $z \approx 3.19$  
characterized by typical low-ionization lines. 
Based on a measurement of the \ion{O}{1}\,$\lambda 1039$ line, 
the authors reported that one of these systems displays 
an enrichment level comparable to the Milky Way interstellar medium
($\gtrsim \frac{1}{2}$ solar). 
The highest redshift system analyzed is the one responsible for 
the obvious LLS feature below $\sim 4200$\AA.  
At $z_{abs} \sim 3.55$, this was at the time the highest 
redshift LLS system studied.  
Limited by spectral resolution, the
authors presented only a cursory analysis of its
properties.    These absorption systems, especially the LLS,
are the focus of the 
current paper\footnote{The fourth system (at $z=3.33$)
noted by \cite{hunstead86} has a poorly constrained \nhi\ value and 
we have chosen not to analyze its properties in detail.}.
Throughout the manuscript we assume 
the solar abundances reported by \cite{gas07}.
%a cosmology consistent with the 5~year
%WMAP results \citep{wmap05} and 

\section{Data}

We observed \qso\ with the MIKE echelle spectrograph  \citep{bernstein03}
on the Magellan II (Clay) Telescope for two 2,400 second exposures on 
the night of September 2, 2004.  
The MIKE spectrometer uses a dichroic centered at $\lambda \approx 5000$\AA\
to separate the optical light into two cameras where it is collimated
and dispersed onto the SITE 2k~x~4k 15$\mu$m detectors.
The data was reduced using the MIKE Reduction 
Pipeline\footnote{http://web.mit.edu/$\sim$burles/www/MIKE/}
\citep{bbp08} which employs novel
techniques for the flat fielding, sky subtraction, spectral extraction, 
wavelength and flux calibration of echelle spectroscopy.  
The reduction algorithms also estimate and subtract scattered light 
detected between the echelle orders.  This signal smoothly
varies across the image and has a negligible contribution compared
to the sky, even at blue wavelengths.

\begin{figure*}
\begin{center}
\includegraphics[height=6.8in,angle=90]{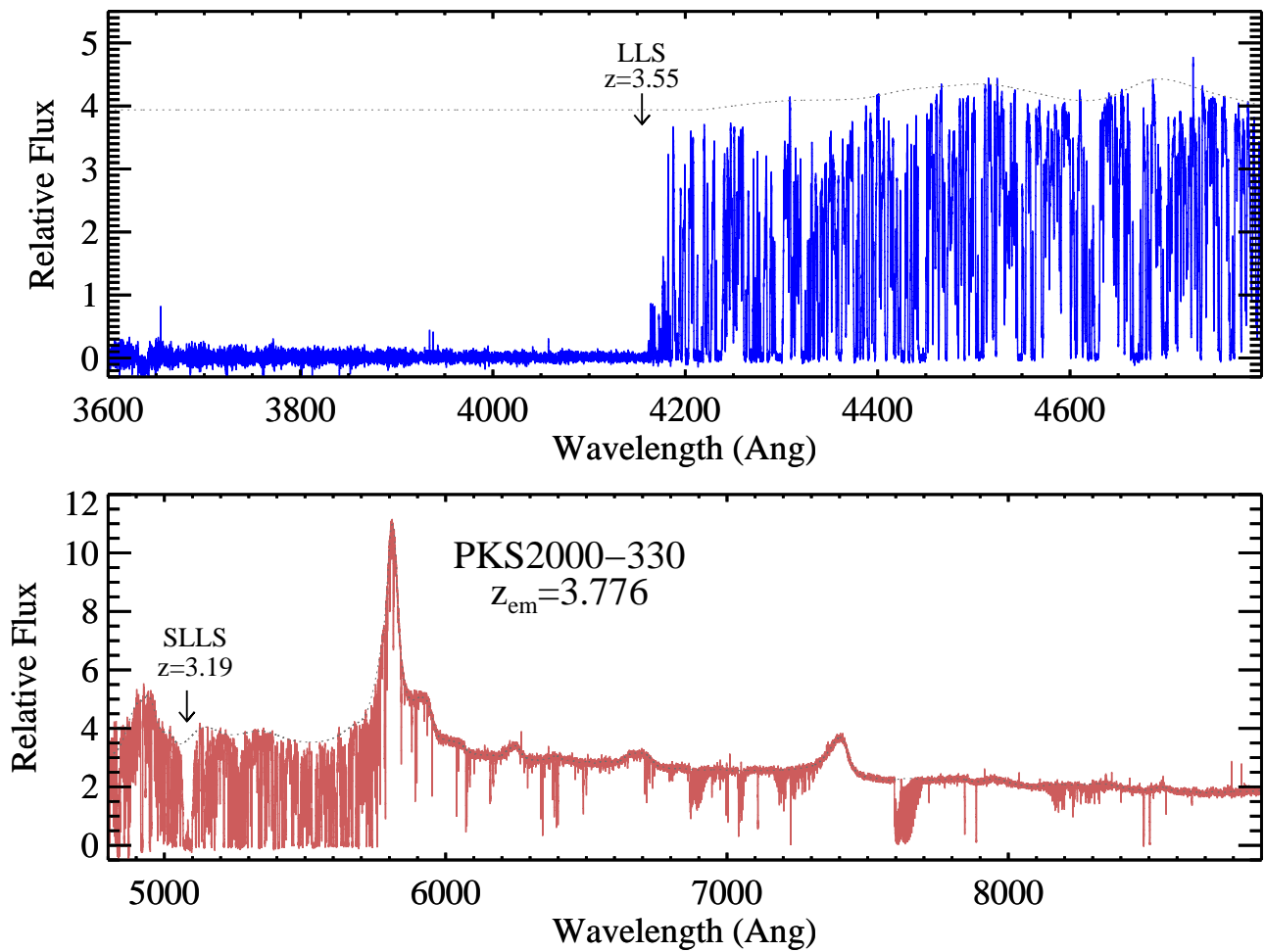}
\end{center}
\caption{Magellan/MIKE spectrum of the quasar \qso\ observed on the night
of September 2, 2004 (4800s total exposure).  The upper panel shows the 
data acquired through the red camera at a dispersion of FWHM$\approx 12 \mkms$.  
Strong \lya\ absorption from a complex
of super-LLS is indicated at $\lambda \approx 5100$\AA.
The lower panel plots the data acquired through the blue camera
at a dispersion of FWHM~$\approx 10 \mkms$.
We mark the Lyman limit system at $\lambda \approx 4100$\AA\
corresponding to $z \approx 3.55$.
The data have been smoothed in each panel by 3~pixels for presentation
purposes only (finer detail cannot be resolved by this figure).
The dotted line in each panel shows our estimate of the quasar continuum 
which is used to normalize the spectra for absorption-line analysis.
  \label{fig:spec}}
\end{figure*}

The resolution of this 
data is FWHM$\sim 10 \mkms$ for the blue data ($3200-5000$~\AA) 
and $\sim 12 \mkms$ for the red data ($5000-9000$~\AA)
as measured  from a series of well-isolated arc lines.
The signal-to-noise ($S/N$) of the data is $\approx 25$ per
3\kms\ and 4.2\kms\ pixel at $\lambda = $4400\AA\ and 6000\AA,
on the blue and red sides respectively.  
Because of the differing spectral resolution, we did not 
coadd the blue and red-side data in the region of spectral
overlap.  Instead, we restrict the analysis of features at 
$\lambda < 5000$\AA\ to the blue-side spectrum and 
$\lambda \ge 5000$\AA\ to the red-side data.

The calibrated spectra are presented in Figure~\ref{fig:spec}.
It is our experience that the relative flux calibration is accurate
to within 15\% \citep{bbp08}.
To analyze the absorption lines in a quasar spectrum,
one must first estimate its intrinsic continuum.  
We fit a series of b-splines and low-order 
polynomials to our spectrum of \qso\
by interactively selecting regions where a visible
inspection suggests minimal absorption,  
i.e.\ the regions were assumed to reflect the quasar continuum.
This process is augmented with corrections made by hand, generally
to maintain a smooth curve.
The continuum estimate is the leading source of
systematic error, especially for analysis within the \lya\ forest.  
At this high redshift, there are few (if any) narrow 
bands of unabsorbed quasar light blueward of the \lya\ emission peak. 
For spectra with S/N typical of our dataset,
it is difficult to avoid a systematic
underestimate of the continuum at these wavelengths.  
This is especially true for data blueward of the rest-frame Lyman limit
for the quasar
($\lambda < 4350$\AA) where absorption from all members of the
Lyman series may contribute.
In Figure~\ref{fig:spec} we show our estimate
of the quasar continuum as a light dotted line.  The reduced
and normalized 1-d spectra are available at 
http://www.ucolick.org/$\sim$xavier/LLS.

\section{Ionic Column Densities}

In this section, we discuss the techniques and measurement of ionic
column densities for the gas comprising the LLS at $z\approx 3.55$
toward \qso.
An analysis of the column densities 
for the SLLSs at $z\approx 3.19$
is presented in Appendix~\ref{sec:appx}.

\subsection{Techniques}

The standard method of characterizing an absorption 
line is with a Voigt profile described by three physical
parameters: the redshift $z$,
the column density $N$, and the Doppler parameter ($b \equiv \sigma \sqrt{2}$).  
These values bear upon the physical quantities that determine the gas
ionization, most importantly the spatial density, kinematics, and temperature.  
To fit for these parameters,  we employed the VPFIT software package
kindly provided by R. Carswell and 
J. Webb\footnote{See http://www.ast.cam.ac.uk/$\sim$rfc/vpfit.html}.  
For this software package, the user supplies
(1) spectral regions for analysis, 
(2) a list of components with ions specified, 
(3) initial guesses for the parameter values, 
and
(4) the spectral line-profile parameterized here as a Gaussian with
FWHM estimated from the ThAr data.
In our analysis, we have generally avoided spectral regions that are
significantly blended with coincident \lya\ forest absorption or 
coincident metal-line absorption. 
We typically have constrained the low-ion\footnote{Low-ions are
defined to be the first ionization state for a gas that has an ionization
potential exceeding 1\,Ryd (e.g.\ O$^0$, Si$^+$, Fe$^+$).  These
atoms and  ions are the dominant ionization state in a neutral hydrogen
gas where a significant far-UV ($h\nu \lesssim 13$\,eV) radiation field
is present.} 
components of a system to 
have identical redshift so that the redshift for each set of low-ion
transitions of a given component are parameterized by a single value.
This constraint is observationally motivated by the good alignment
between the absorption lines in velocity space.
For \ion{H}{1} Lyman lines, which suffer from significant line-blending, 
we have anchored their redshift to the value derived
from an independent fit of the metal-line transitions. 
This presumes that the \ion{H}{1} absorption traces the
low-ion profiles in velocity space, but the
relative column densities are allowed to vary freely from
component to component.  
Unfortunately, in the LLS presented 
here the line-blending is too severe to test the former
assumption empirically.
Future analysis of the other LLS in our survey will address this point
although we note that such analysis is best performed in the low redshift
universe where line-blending with the \lya\ forest, is minimized
\citep[e.g.][]{lpk+09}.
For lines analyzed within the \lya\ forest we allow for a
systematic uncertainty in the continuum placement of 10\%.
This error does not dominate in any of the analysis presented
in this paper.  
In general, our fits adopt the fewest components which
yield reduced $\chi^2$ values near unity.

We complement this line-profile analysis with
ionic column densities measured from the apparent optical
depth method \citep[AODM;][]{ss96}.  In the AODM, one
converts the observed normalized flux profile ($f$) to an apparent
optical depth array $\tau_a \equiv -\ln(1/f)$ and then sums
over the velocity interval of the line-profile 
to calculate the integrated column density.
This non-parametric technique provides accurate results 
for well-resolved profiles. 
It also provides conservative upper and lower limits for 
undetected or saturated lines.  
Throughout this paper, we adopt these limits where appropriate.
Upper limits correspond to $2\sigma$ statistical uncertainties and
lower limits correspond to the column densities by demanding
$f \ge 1\sigma$ in every pixel.
Finally, the AODM analysis is useful for
identifying unresolved components (hidden saturation) by comparing
the column densities of a series of transitions with different
oscillator strengths from a single ion.
This analysis is especially appropriate for our dataset because the spectral
resolution is lower than the echelle data commonly 
used in quasar absorption line studies.
For the atomic data, we have relied on the compilation of atomic
data given by \cite{morton03}.

\subsection{The $z \approx 3.55$ LLS Intervening \qso}
\label{sec:z355}

%\subsubsection{LLS at $z=3.55$}

The absorption system at $z\approx 3.55$ gives rise to the strong Lyman limit 
observed in the spectrum of \qso\ 
(Figure~\ref{fig:spec}) and, therefore, 
is the only LLS showing the complete Lyman series. 
As demonstrated below, 
spectral coverage that includes the entire Lyman series
is crucial to precisely constraining the 
\ion{H}{1} column density. 
We place a lower limit on the total \ion{H}{1} column
density of this LLS by estimating an upper limit to
the normalized flux
shortward of the Lyman limit.  In the spectral window
$\lambda = 4000-4050$\AA, we measure an upper limit to the
normalized flux $f_{LL} < 0.004$ (95$\%$ c.l.) where we have
conservatively allowed for a 50\% lower continuum then presented in 
Figure~\ref{fig:spec}.  This implies a lower limit to the
total \ion{H}{1} column density of the LLS,
$\mnhi > \tau_{LL} / \sigma(E) = 10^{17.7} \cm{-2}$
where $\tau_{LL} = \ln(1/f_{LL})$ and $\sigma(E)$ is 
evaluated at $E = {\rm 4025 \AA / [911.76 \AA (1+z_{LL})]}$\,Ryd.
This assumes that the contribution of coincident absorption
by the IGM and metal-lines is minor, which is reasonable 
for wavelengths longward of the next lower redshift LLSs.
We have also ruled out coincident, strong \lya\ absorption at
these wavelengths by searching for corresponding metal-line
absorption (e.g.\ \ion{C}{4}).

\begin{figure}
\begin{center}
\includegraphics[height=3.5in,angle=90]{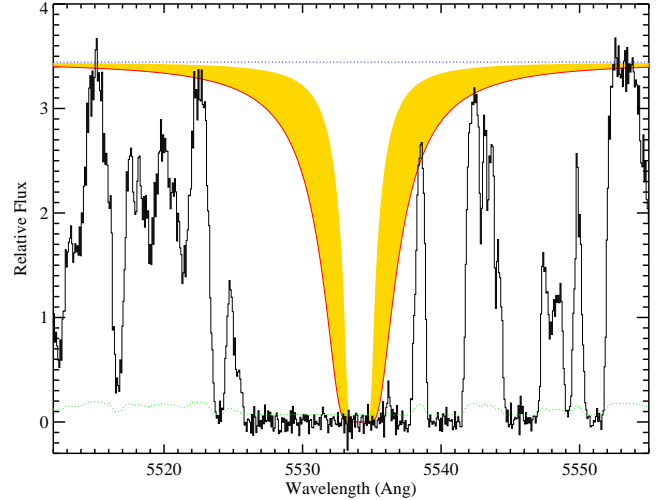}
\end{center}
\caption{
Spectral region encompassing the \lya\ transition of the 
LLS at $z \approx 3.55$. The dotted
line at relative flux $f \approx 3.44$ traces the continuum
estimated from assumed regions with minimal absorption within
and nearby this spectral region.  Overplotted on the data
is the set of \lya\ models corresponding to $\mnhi = 10^{18.1}$ to
$10^{18.85} \cm{-2}$ (solid curve) for a single absorber centered
at $z=3.5523$, where the majority of \ion{H}{1} gas is expected
to lie in this LLS (see below).  The profile assuming 
$\mnhi = 10^{18.85} \cm{-2}$ significantly underpredicts
the flux at $\lambda \approx 5538.5$\AA.  Given that the flux
at these wavelengths is further depressed by the `clouds'
centered at $\lambda \approx 5537$\AA\ and 5540\AA, we adopt
a conservative upper limit to the \nhi\ value at $z=3.5523$
of $10^{18.85} \cm{-2}$.
  \label{fig:Lya}}
\end{figure}

We can set a conservative upper limit on the total \ion{H}{1}
column density through analysis of the \lya\ transition.
Figure~\ref{fig:Lya} presents the spectrum of covering the 
\lya\ transition for the $z \approx 3.55$ LLS.  
The dotted
line at relative flux $f \approx 3.44$ traces the continuum
estimated from regions of minimal absorption within
and nearby this spectral region.  Overplotted on the data
is the set of \lya\ models corresponding to $\mnhi = 10^{18.1}$ to
$10^{18.85} \cm{-2}$ for a single absorber centered
at $z=3.5523$, where the majority of \ion{H}{1} gas is expected
to lie (see below).  The profile assuming 
$\mnhi = 10^{18.85} \cm{-2}$ significantly underpredicts
the flux at $\lambda \approx 5538.5$\AA, especially if one accounts
for the additional \lya\ absorption from the clouds at  
$\lambda \approx 5537$\AA\ and 5540\AA.  We adopt
a conservative upper limit for the gas at $z=3.5523$
of $\mnhi \le 10^{18.85} \cm{-2}$.

A line-profile analysis of the full Lyman series will, in principle,
give a more precise constraint on the \ion{H}{1} column density of the LLS.
Furthermore, such an analysis may also resolve the \ion{H}{1} system
into multiple velocity components.
There are several challenges, however, to this analysis:
(i) the various components of the LLS may blend with one another, 
especially in the higher-order Lyman series transitions;
(ii) the Lyman series of the LLS is severely blended with the
lower redshift IGM;
and
(iii) the majority of lines lie on the flat portion of the curve-of-growth
implying a significant degeneracy between $b$ and \nhi.
We address the first concern (in part) by demanding that the strongest
\ion{H}{1} clouds have identical redshift as the strongest low-ion absorption.
This requirement follows from the 
assumption that low-ions primarily arise in regions of
large surface density of \ion{H}{1} gas that can self-shield
the material from ionizing photons.
We address the second point by avoiding severely blended lines.
On the third point, we consider physically plausible values for the 
Doppler parameter and adopt the corresponding \nhi\ constraints.

\begin{figure*}
\begin{center}
\includegraphics[height=6.8in]{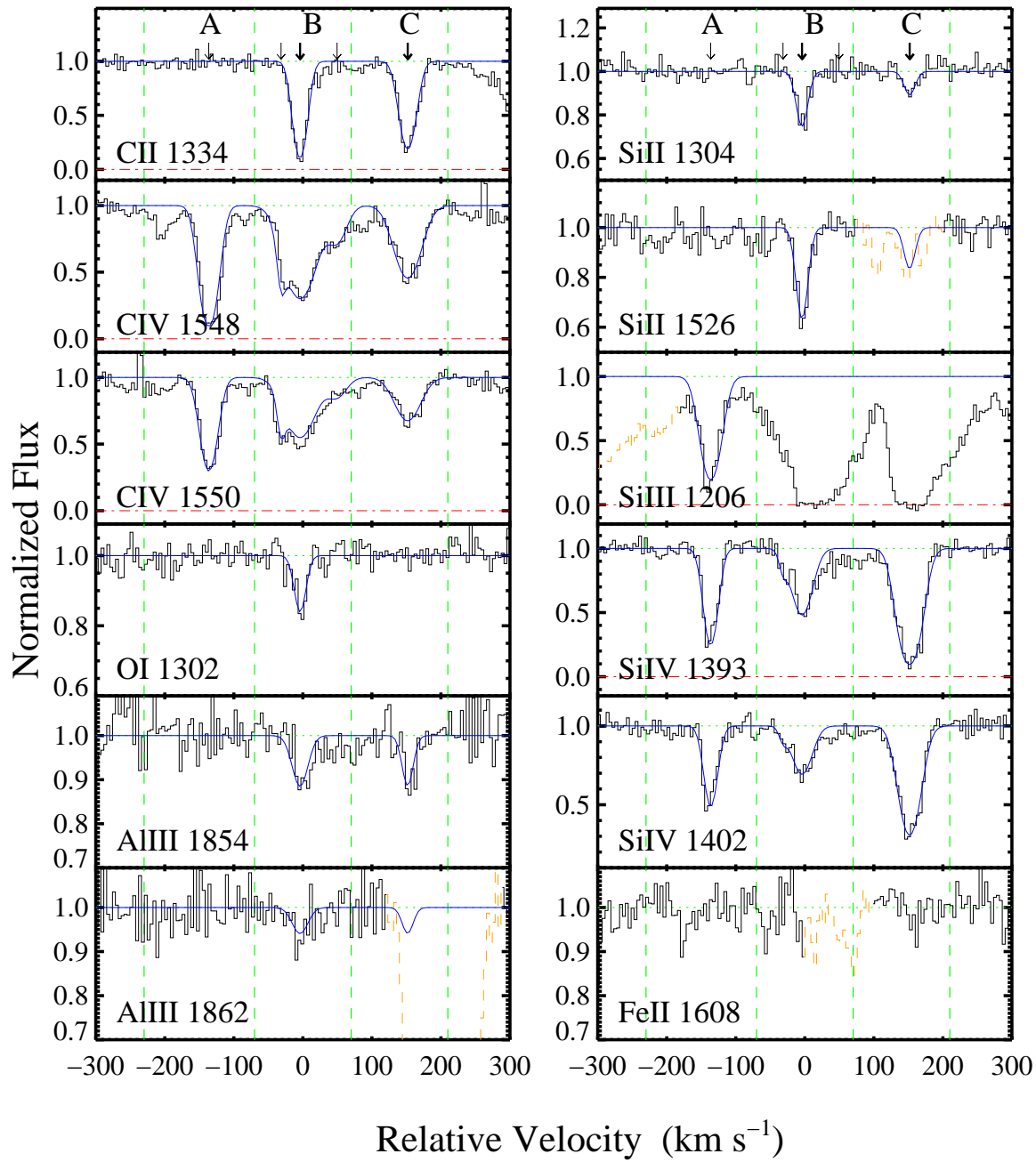}
\end{center}
\caption{
Velocity plot of the principal metal-line transitions detected
for the LLS at $z \approx 3.55$ ($v=0\mkms$ corresponds to $z=3.550$).  
In each panel, the vertical (green)
dashed lines designate three subsystems (A,B,C) that we have designated.
Blends with coincident absorption (e.g.\ \lya\ lines) are denoted as
(orange) dashed spectra.  
Overplotted on the data is the best-fit
solution from a line-profile analysis of the data using the VPFIT
software package.  Arrows in the top panels locate the 
centroids of the velocity components (Table~\ref{tab:llsvpfit}).
  \label{fig:z355mtl}}
\end{figure*}

\begin{deluxetable*}{cccccccccc}
\tablewidth{0pc}
\tablecaption{VPFIT SOLUTIONS FOR THE LLS AT $z=3.550$\label{tab:llsvpfit}}
\tabletypesize{\footnotesize}
\tablehead{\colhead{Subsystem} & \colhead{$z$} & \colhead{$\sigma(z)$} 
& \colhead{$v^a$} & \colhead{Ion} &
\colhead{$b$} & \colhead{$\sigma(b)$}
& \colhead{$\log N$} & \colhead{$\sigma(N)$} \\
& & ($10^{-5}$)& (\kms) && (\kms) & (\kms)  }
\startdata
A&3.547932&0.5&$-136$&\ion{H}{1}&17.45& 3.27&14.88& 0.10\\
&&&&\ion{C}{4}&14.09& 0.60&13.95& 0.03\\
&&&&\ion{Si}{4}&10.90& 1.34&13.24& 0.04\\
&&&&\ion{Si}{3}&16.68& 1.32&13.00& 0.03\\
B&3.549521&1.1&$ -32$&\ion{C}{4}& 3.00& 1.56&13.23& 0.16\\
&&&&\ion{Si}{4}& 3.00& 9.69&12.09& 0.17\\
B&3.549938&0.5&$  -4$&\ion{H}{1}&19.38& 4.53&16.82& 0.20\\
&&&&\ion{C}{4}&27.48& 1.53&13.89& 0.03\\
&&&&\ion{C}{2}&10.01& 0.64&14.09& 0.04\\
&&&&\ion{O}{1}& 9.24& 3.78&13.34& 0.09\\
&&&&\ion{Al}{3}&14.10& 8.28&12.11& 0.17\\
&&&&\ion{Si}{2}& 8.97& 1.50&13.27& 0.04\\
&&&&\ion{Si}{4}&17.97& 1.53&13.11& 0.03\\
C&3.550751&3.8&$+  50$&\ion{C}{4}&17.49& 4.60&13.13& 0.09\\
C&3.552304&0.4&$+ 152$&\ion{H}{1}&25.00& 2.50&18.00& 0.25\\
&&&&\ion{C}{4}&23.71& 1.11&13.65& 0.03\\
&&&&\ion{C}{2}&12.86& 0.42&14.00& 0.03\\
&&&&\ion{Al}{3}& 8.58& 0.42&11.96& 0.18\\
&&&&\ion{Si}{4}&18.67& 0.48&13.64& 0.03\\
&&&&\ion{Si}{2}& 8.41& 0.42&12.84& 0.11\\
\enddata
\tablenotetext{a}{Velocity relative to $z=3.550$.}
\end{deluxetable*}

In Figure~\ref{fig:z355mtl},
we present a series of metal-line transitions for the LLS at 
$z \approx 3.55$.  We have divided the line-profiles into 
three subsystems:  (A) $-230 \mkms \le \delta v < -70 \mkms$, 
(B) $-70 \mkms \le \delta v < 70 \mkms$, and
(C) $70 \mkms \le \delta v < 210 \mkms$ relative to
$z_{LLS} = 3.550$.
The velocity interval of the subsystems was chosen to separate 
the two significant
low-ion complexes from each other (B and C) and also to separate
strong \ion{C}{4} absorption (A) from the other subsystems. 
Subsystems~B and C show significant low-ion absorption and
we expect that these contain the majority of the \ion{H}{1} gas.
We have independently fit the metal-line profiles of the various subsystems
using the VPFIT software package.  We have forced the
components comprising each subsystem to have identical redshift
for different low-ions but have allowed the Doppler parameter values
to vary because the line widths of
the high-ion transitions appear systematically wider 
than those of the low-ions.
The best-fit solutions and $1\sigma$ error estimates are presented
in Table~\ref{tab:llsvpfit} and the models are overplotted on the
data in Figure~\ref{fig:z355mtl}.
The low and high-ion absorption in each subsystem is well modeled
by components having identical velocity, but the analysis 
also suggests
that the high-ion gas has systematically larger Doppler
parameters ($b \gtrsim 20 \mkms$) than the low-ion gas ($b \lesssim 15 \mkms$).
This conclusion, however, is sensitive to the details of the component
structure assumed in the analysis.  It is possible to achieve an acceptable
model where the $b$-values are identical between the low and high-ions,
but this would require the inclusions of yet further components.
The profile model presented here is the simplest one that reproduces
the observations.
We return to this point when discussing photoionization models
of the system ($\S$~\ref{sec:photo}).
Finally, we note that the regions containing the transitions of 
the \ion{O}{6} and \ion{N}{5} doublets
are hopelessly blended with coincident IGM features.

\begin{figure*}
\begin{center}
\includegraphics[height=6.8in]{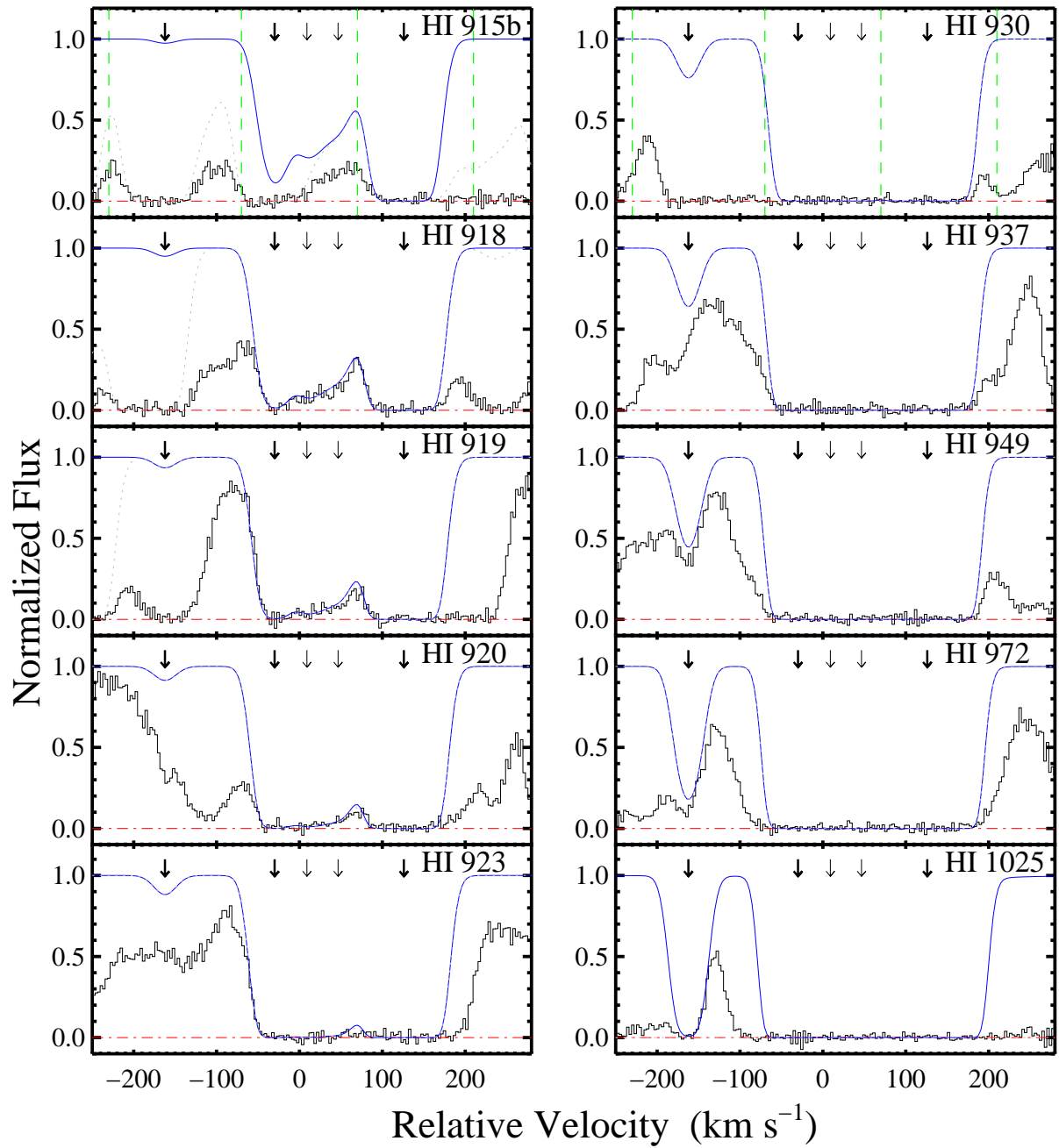}
\end{center}
\caption{Velocity plots of the Lyman limit (upper-left panel) and 
a set of \ion{H}{1} Lyman series transitions for the LLS at
$z \approx 3.55$ toward \qso.  Arrows designate the centroids of
the various velocity components used in the analysis (Table~\ref{tab:llsvpfit}).
The solid blue curve shows the convolved line-profile for each Lyman line
individually, while the dotted gray curve plots the convolved solution for
all Lyman lines (inspect the panel showing the 
\ion{H}{1}~918 transition where multiple Lyman lines overlap).
The vertical dashed lines in the panels showing \ion{H}{1}~915b and 930
designate subsystems A, B, and C.  The velocity $v=0\mkms$ corresponds
to an arbitrary $z=3.550$.
  \label{fig:z355hi}}
\end{figure*}

The redshifts of the strongest low-ion component in each of the subsystems
were then taken as an input constraint for the \ion{H}{1} Lyman 
series analysis.
In Figure~\ref{fig:z355hi}, we present a subset of the 
Lyman series transitions and also a portion of the data near
the Lyman limit.   It is evident that the strongest, highly saturated
members of the Lyman series offer only a weak constraint 
on the \ion{H}{1} column
densities of the subsystems.
%analysis proceeded in an iterative, semi-quantitative fashion.
We first visually identified spectral regions that were most likely to 
constrain the strongest \ion{H}{1} components of subsystems A, B, and C.
We then inputted to VPFIT these regions, a set of \ion{H}{1} absorption lines
at the redshifts of each subsystem 
($z_A = 3.547932, z_B = 3.549938, z_C=3.552304$), and 
additional \ion{H}{1} lines within subsystem~B (with $z_C > z > z_B$) to model
additional absorption evident in the Lyman series. 
The non-zero flux
at $\delta v \approx 50\mkms$ in the high-order Lyman series lines
(e.g.\ \ion{H}{1} 918, 919) indicates that the integrated column densities
of these additional components is significantly lower than the
\ion{H}{1} column density of subsystem~C and possibly subsystem~B.
We can set a conservative upper limit to the total \nhi\ of this
gas by integrating the apparent optical depth of the \ion{H}{1}~918
profile in the interval $\delta v = +17$ to $+80 \mkms$,
finding $\mnhi < 10^{16.9} \cm{-2}$ (95$\%$ c.l.).
This value is to be considered an upper limit because there may be
significant line-blending with lines from lower redshift IGM systems. 

Our trials with the VPFIT package yielded a set of models 
with acceptable reduced $\chi^2$ values, but
we found that the parameters of the additional components in subsystem~B
are very poorly constrained and that we could
not achieve a unique solution for the full set of profiles.
Therefore, we used the range of solutions to guide a `by-eye' fitting
analysis of the components at $z = z_B$ and $z_C$.  
The parameters for the gas at $z = z_B$ are well constrained
by the higher order Lyman series (\ion{H}{1}~918,919,920,923):
$\log \mnhi = 16.8 \pm 0.2$, $b_{\rm HI}^B = 19.4 \pm 4.5 \mkms$.
There is a weak dependence of these values on our assumed
parameters for absorption
at $\delta v \approx +20 \mkms$ which contributes to our error estimate.  
Nevertheless, the total \nhi\ value for the gas in subsystem~B
is much less than that required to explain the observed absorption
at $\lambda < 4150$\AA.  
We conclude that the majority of \ion{H}{1} gas in the LLS
at $z \approx 3.55$ is associated with subsystem~C.

The detection of positive flux at $\delta v \approx +180 \mkms$
in \ion{H}{1}~937 and at $\delta v \approx +80\mkms$ in \ion{H}{1}~918,919
sets an upper limit to the combined \nhi\ and $b_{\rm HI}^C$ values of an
\ion{H}{1} component centered at $z=z_C$.  
Formally, the data permit \nhi\ values as large as $10^{18.85} \cm{-2}$
provided $b$-values less than 10\kms.  We adopt a prior of
$b_{\rm HI}^C \ge 20 \mkms$, however, based on the following physical
arguments.  First, we demonstrate in the following section that
the gas related to subsystem~C is highly ionized.  Even if we 
assume a soft (i.e.\ stellar) ionizing radiation field,
the implied gas temperature is $T > 15,000$K giving $b > 16\mkms$.
Second, systems with $b \le 20 \mkms$ are extremely rare, as
evidenced by the rarity of D/H measurements from the IGM.
Third, the \ion{C}{2} transitions have Doppler parameters
$b > 10 \mkms$ which set a lower limit to the Doppler parameter
of the \ion{H}{1} gas.  Finally, a model with $b \le 20 \mkms$
would underpredict the absorption at 
$\delta v \approx -45 \mkms$ and $+60\mkms$
for the entire Lyman series.  Therefore, one would need to introduce
an additional 10 or more absorption lines (including many at unrelated
redshifts) to reproduce the observations.  
Together, these points motivate a lower limit to $b_{\rm HI}^C$
of $\approx 20 \mkms$.

%Assuming only minor absorption
%from neighboring (or coincident) lines at these positions, we derive
%an upper limit to the \ion{H}{1} column density for $b \ge 20 \mkms$ of
%$\log \mnhi < 18.3$.  Formally, the data permit larger \nhi\ values
%for lower $b$ values but we expect $b \ge 20\mkms$ because 
%(i) the \ion{H}{1} lines of the intergalactic medium rarely have 
%$b < 20 \mkms$ \citep[][but note that this has not been established
%for the LLSs]{kt97}; 
%(ii) the metal absorption for subsystem~C is well modeled by 
%a single component with $b > 15 \mkms$; 
%(iii) one would need to introduce additional absorption lines
%at $\delta v \approx -45 \mkms$ and $+60\mkms$
%for {\it all} of the Lyman series transitions to fit the profiles
%with such a small $b$-value for component~C.
%The simplest hypothesis is to assume a single \ion{H}{1} component
%with $b \ge 20 \mkms$.

With this prior on $b_{\rm HI}^C$, the data require $\mnhi < 18.5 \cm{-2}$ and we
adopt a best-estimate of $\mnhi = 10^{18.0 \pm 0.25} \cm{-2}$
and $b_{\rm HI}^C = 25 \pm 2.5 \mkms$ for subsystem~C.
We caution that subsequent analysis should not treat the \nhi\ or $b$-values
for subsystem~C as following a normal distribution with the reported
uncertainties.   Instead, we recommend that the central value be 
considered uniformly distributed.  But we also note that the 
\nhi\ value cannot be significantly less than $10^{17.7} \cm{-2}$
because of the combined constraints from the Lyman limit and subsystem~B.

\begin{deluxetable}{cccccccccc}
\tablewidth{0pc}
\tablecaption{CONSTRAINTS ON \nhi\ FOR THE LLS AT $z=3.550$\label{tab:HIconstraints}}
\tabletypesize{\footnotesize}
\tablehead{\colhead{Subsystem} & \colhead{Feature} & \colhead{$\log \mnhi$} 
& \colhead{Assumptions} }
\startdata
A+B+C  & Lyman Limit      & $> 17.7$       & Minimal IGM blending \\
C      & \lya\            & $<18.85$       & None \\
A      & \lyb, \lyg, \lyd & $15.0 \pm 0.1$ & Minimal IGM blending \\
B      & Ly8-11           & $16.8 \pm 0.2$ & Minimal IGM blending \\
C      & Ly5, Ly10-11     & $18.0 \pm 0.25$& $b \ge 20 \mkms$     \\
\enddata
%\tablenotetext{a}{Velocity relative to $z=3.550393$.}
\end{deluxetable}

\begin{deluxetable*}{lccccccccc}
\tablewidth{0pc}
\tablecaption{IONIC COLUMN DENSITIES FOR THE LLS AT $z=3.55$\label{tab:llsclm}}
\tabletypesize{\footnotesize}
\tablehead{\colhead{Ion} & \colhead{$\lambda_{\rm rest}$} & \colhead{$\log f$}
& \colhead{$v_{int}^a$} 
& \colhead{$\log N_{\rm AODM}$}&
 \colhead{$\log N_{\rm VPFIT}$}&
  \colhead{$\log N_{\rm adopt}$} \\
& (\AA) & & (\kms) &  }
\startdata
\cutinhead{SUBSYSTEM A}
\ion{H}{1} & 1215.6701 & $ -0.3805$&&&&$14.88\pm 0.20$\\
\ion{C}{2}&1334.5323 &$ -0.8935$&$[ -186,  -86]$&$< 12.75$&&$< 12.75$\\
\ion{C}{4}&1548.1950 &$ -0.7194$&$[ -186,  -96]$&$ 13.91 \pm 0.03$&$ 13.95 \pm 0.03$&$ 13.95 \pm 0.03$\\
&1550.7700 &$ -1.0213$&$[ -186,  -96]$&$ 13.93 \pm 0.03$&&\\
\ion{O}{1}&1302.1685 &$ -1.3110$&$[ -166,  -96]$&$< 13.00$&&$< 13.00$\\
\ion{Al}{3}&1854.7164 &$ -0.2684$&$[ -186,  -86]$&$< 12.12$&&$< 12.12$\\
&1862.7895 &$ -0.5719$&$[ -186,  -86]$&$< 12.37$&&\\
\ion{Si}{2}&1304.3702 &$ -1.0269$&$[ -186,  -86]$&$< 12.80$&&$< 12.72$\\
&1526.7066 &$ -0.8962$&$[ -166,  -96]$&$< 12.72$&&\\
\ion{Si}{3}&&&&&$ 13.00 \pm 0.03$&$ 13.00 \pm 0.03$\\
\ion{Si}{4}&1393.7550 &$ -0.2774$&$[ -186,  -96]$&$ 13.14 \pm 0.03$&$ 13.24 \pm 0.04$&$ 13.24 \pm 0.04$\\
&1402.7700 &$ -0.5817$&$[ -186,  -96]$&$ 13.23 \pm 0.03$&&\\
\ion{Fe}{2}&1608.4511 &$ -1.2366$&$[ -186,  -86]$&$< 12.93$&&$< 12.93$\\
\cutinhead{SUBSYSTEM B}
\ion{H}{1} & 1215.6701 & $ -0.3805$&&&&$16.82\pm 0.20$\\
\ion{C}{2}&1036.3367 &$ -0.9097$&$[  -47,   33]$&$ 14.12 \pm 0.03$&$ 14.09 \pm 0.04$&$ 14.09 \pm 0.04$\\
&1334.5323 &$ -0.8935$&$[  -57,   43]$&$ 14.03 \pm 0.03$&&\\
&1335.7077 &$ -0.9397$&$[  -47,   23]$&$< 13.36$&&\\
\ion{C}{4}&1548.1950 &$ -0.7194$&$[  -57,   33]$&$ 13.89 \pm 0.03$&$ 14.03 \pm 0.03$&$ 14.03 \pm 0.03$\\
&1550.7700 &$ -1.0213$&$[  -57,   33]$&$ 14.05 \pm 0.03$&&\\
\ion{O}{1}&1302.1685 &$ -1.3110$&$[  -37,   33]$&$ 13.35 \pm 0.07$&$ 13.34 \pm 0.09$&$ 13.34 \pm 0.09$\\
\ion{Al}{3}&1854.7164 &$ -0.2684$&$[  -57,   43]$&$ 12.19 \pm 0.13$&$ 12.11 \pm 0.17$&$ 12.11 \pm 0.17$\\
&1862.7895 &$ -0.5719$&$[  -57,   43]$&$< 12.38$&&\\
\ion{Si}{2}&1304.3702 &$ -1.0269$&$[  -57,   43]$&$ 13.21 \pm 0.06$&$ 13.27 \pm 0.04$&$ 13.27 \pm 0.04$\\
&1526.7066 &$ -0.8962$&$[  -37,   33]$&$ 13.29 \pm 0.03$&&\\
\ion{Si}{4}&1393.7550 &$ -0.2774$&$[  -57,   33]$&$ 13.14 \pm 0.03$&$ 13.15 \pm 0.03$&$ 13.15 \pm 0.03$\\
&1402.7700 &$ -0.5817$&$[  -57,   33]$&$ 13.18 \pm 0.03$&&\\
\ion{Fe}{2}&1608.4511 &$ -1.2366$&$[  -57,   43]$&$< 13.09$&&$< 13.09$\\
\cutinhead{SUBSYSTEM C}
\ion{H}{1} & 1215.6701 & $ -0.3805$&&&&$18.00\pm 0.25$\\
\ion{C}{2}&1036.3367 &$ -0.9097$&$[  112,  192]$&$ 14.20 \pm 0.03$&$ 14.00 \pm 0.03$&$ 14.00 \pm 0.03$\\
&1334.5323 &$ -0.8935$&$[  102,  202]$&$ 13.97 \pm 0.03$&&\\
&1335.7077 &$ -0.9397$&$[  122,  192]$&$ 12.90 \pm 0.10$&&\\
\ion{C}{4}&1548.1950 &$ -0.7194$&$[  102,  192]$&$ 13.65 \pm 0.03$&$ 13.65 \pm 0.03$&$ 13.65 \pm 0.03$\\
&1550.7700 &$ -1.0213$&$[  102,  192]$&$ 13.64 \pm 0.03$&&\\
\ion{O}{1}&1302.1685 &$ -1.3110$&$[  122,  192]$&$< 13.01$&&$< 13.01$\\
\ion{Al}{3}&1854.7164 &$ -0.2684$&$[  102,  202]$&$< 12.13$&$ 11.96 \pm 0.18$&$ 11.96 \pm 0.18$\\
&1862.7895 &$ -0.5719$&$[  102,  202]$&$< 14.04$&&\\
\ion{Si}{2}&1304.3702 &$ -1.0269$&$[  112,  192]$&$ 12.79 \pm 0.14$&$ 12.84 \pm 0.11$&$ 12.84 \pm 0.11$\\
\ion{Si}{3}&1206.5000 &$  0.2201$&$[  102,  202]$&$> 13.59$&&$> 13.59$\\
\ion{Si}{4}&1393.7550 &$ -0.2774$&$[  102,  192]$&$ 13.60 \pm 0.03$&$ 13.64 \pm 0.03$&$ 13.64 \pm 0.03$\\
&1402.7700 &$ -0.5817$&$[  102,  192]$&$ 13.62 \pm 0.03$&&\\
\ion{Fe}{2}&1608.4511 &$ -1.2366$&$[  102,  202]$&$< 12.91$&&$< 12.91$\\
\enddata
\tablenotetext{a}{Velocity interval for the AODM
relative to $z=3.550000$.}
\end{deluxetable*}

\begin{figure*}
\begin{center}
\includegraphics[height=6.8in]{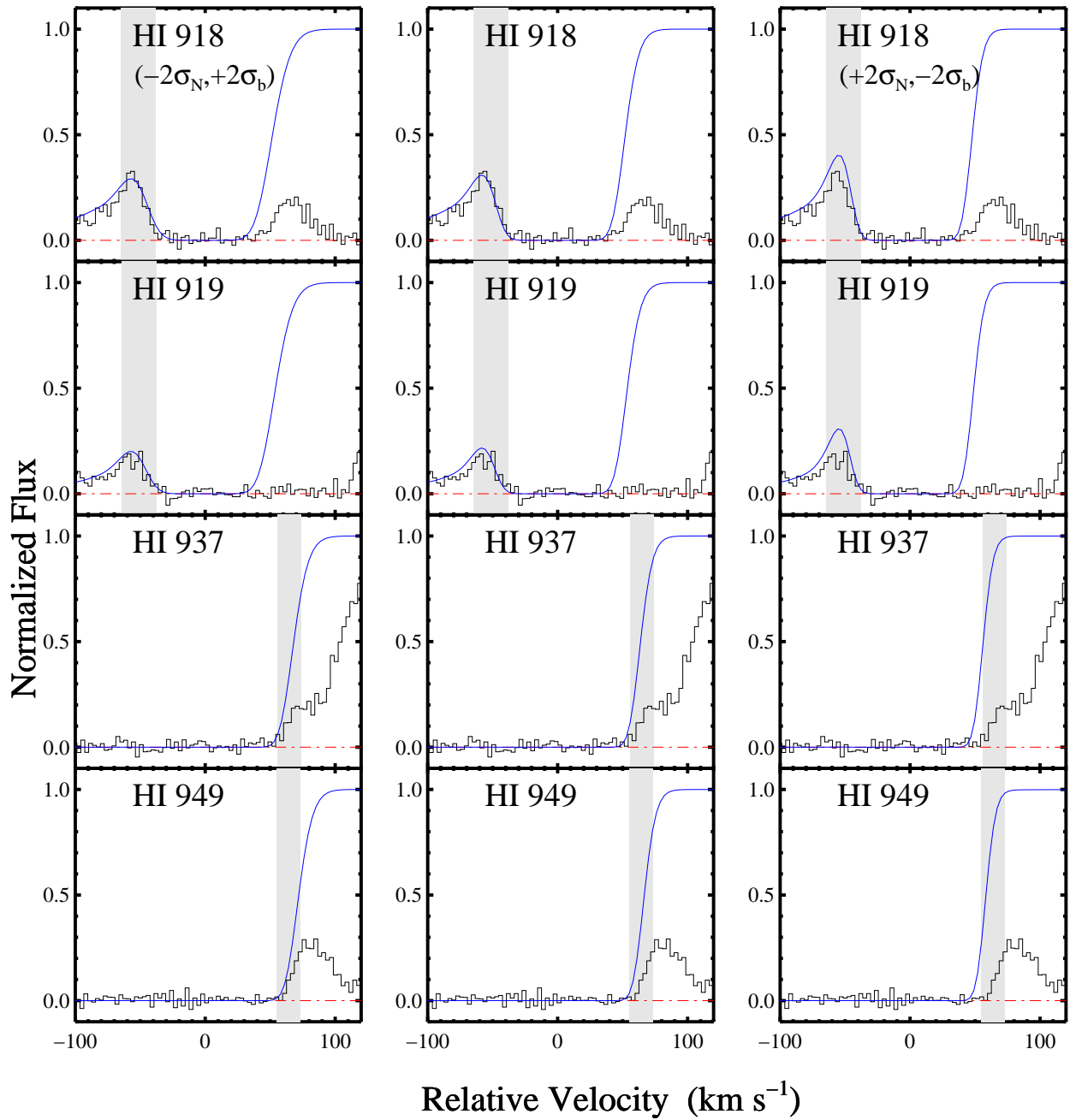}
\end{center}
\caption{
Zoom-in on the Lyman series lines that most tightly constrain
the column densities and $b$-values for the \ion{H}{1} components
comprising subsystem C.  The center panels show the favored model
(solid blue curve) line for the central values of our line-profile solution
while the left/right panels show 2$\sigma$ departures from the
central value chosen to minimize changes in the model, i.e.\ an increase
in \nhi\ is offset by a decrease in the $b$-value.   The shaded 
boxes show the portion of the data that has greatest constraint
on the analysis.
  \label{fig:fiddle}}
\end{figure*}

The full set of constraints on the \nhi\ values for the subsystems
comprising this LLS is summarized in Table~\ref{tab:HIconstraints}.
Our favored solution is overplotted
on the data in Figure~\ref{fig:z355hi}.  It is evident that the
strongest members of the Lyman series ($\alpha, \beta, \gamma, \delta$)
offer little constraint on the \nhi\ values of subsystems B and C.  
The best constraints come from higher order lines;  these are
presented in Figure~\ref{fig:fiddle}. 
In this figure, we present the favored solution (middle panel)
and $2 \sigma$ departures in the \nhi\ and $b$ values 
(incremented in opposition to minimize the change to the model).  The shaded 
regions identify the
pixels that most constrain these parameters.  
A few points should be emphasized.  First, the left-hand panels
correspond to a model with $\mnhi = 10^{17.5} \cm{-2}$ which is
ruled out by observations of the Lyman limit (see above).  
Second, the model in the right-hand panel significantly underpredicts
the absorption at $\delta v \approx +65 \mkms$.
Full coverage of the Lyman series and the Lyman continuum region
at a relatively high S/N ratio has
constrained the \nhi\ value of this LLS.
Such analysis demands echelle spectra of the full Lyman series can
minimize blending
with the $z \sim 3$ IGM and the foreground LLSs.
The full set of ionic column densities for the three subsystems
is provided in Table~\ref{tab:llsclm}.

Before concluding our discussion on the \ion{H}{1} absorption of the LLS at 
$z \approx 3.55$, we remark that our model is not a good
fit to the spectrum at $\lambda \approx 4160-4175$\AA. 
In this spectral range, which corresponds
to the Ly-12, Ly-13, and Ly-14 transitions of the LLS,  
our model underpredicts the absorption at velocities in between
the main components (i.e. away from the line centers).  
This implies either an additional
source of nearly continuous opacity unrelated to the LLS at $z=3.55$ or that we 
have significantly overestimated the quasar continuum flux at 
these wavelengths.
Comparing to the published spectrum of \cite{btt90}, we note a similar
drop in the quasar flux at these wavelengths which argues against the latter
explanation.  Our expectation, therefore,
is that the majority of additional opacity is from unidentified \ion{H}{1}
lines from the intergalactic medium at $z<3.55$.
In any case, the measurements of \nhi\ for this LLS
are well constrained by the Lyman lines redward of 4180\AA.

\section{Photoionization Modeling} 
\label{sec:photo}

At $z>2$, gas clouds with \ion{H}{1} column densities of less 
than $10^{19} \cm{-2}$
are predicted to be photoionized by the extragalactic ultra-violet
background (EUVB) radiation field unless one assumes an extraordinarily
high volume density \citep[e.g.][]{viegas95}.  
In this photoionized gas, the ions observed (especially the low-ions and \ion{H}{1}) 
may represent only a trace quantity of the total gas present.  To determine
the gas abundance and also assess the physical conditions, 
it is necessary to model the ionization state of the gas.
We have performed this modeling for the LLSs
along the sightline
to \qso\ using the Cloudy software package \citep{cloudy98}.  
In practice, we have
calculated ionization models for a series of plane-parallel slabs with
constant gas density $n_H$ and with \nhi\ constrained to match the central values
derived from our line-profile analyses.  We have adopted the EUVB radiation
field calculated by Haardt \& Madau (in prep.; a.k.a.\ CUBA) assuming 
contributions from both quasars and UV-bright galaxies.  We run models
with a wide range of ionization parameters $U \equiv \Phi/n_H c$,
which is akin to adopting an intensity $\Phi$ for the EUVB\footnote{
This intensity has been poorly constrained by  empirical observation,
but see \cite{dww08}.} 
and varying the Hydrogen volume density $n_H$.
Finally, we compare observed ionic ratios against
the model predictions to constrain the ionization parameter $U$
under the assumptions of our simplistic photoionization model.
Ideally, we consider only pairs of ions from the same element to minimize
dependence on the assumed intrinsic abundances.
But we also find valuable constraints from ions of different
elements, even allowing for significant departures from non-solar
relative abundances.
We caution that the single-phase models considered here are overly
simplistic.  In Appendix~\ref{sec:2phase}, we present two-phase
models to explore the impact of more complex scenarios.

As noted above, an absorption system with $\mnhi \ll 10^{20} \cm{-2}$
is predicted to be highly ionized at $z>3$ by the EUVB radiation
field alone.
One signature of significant photoionization is the presence of strong 
high-ion absorption
(\ion{C}{4}, \ion{Si}{4}) relative to low-ion species.  
Indeed, we find
that the ratios of C$^+$/C$^{+3}$ and Si$^+$/Si$^{+3}$ are of order
unity for subsystems~B and C and much less than one for subsystem~A.
Under the assumption of a single-phase model,
these observations indicate the gas in this LLS is predominantly ionized.
As noted in $\S$~\ref{sec:z355}, however, the high-ion gas may have
systematically larger Doppler parameters than the low-ion species.
This might indicate that the 
gas occurs in a distinct region, 
physically separated by a large distance from the low-ion gas.  
We consider this hypothesis improbable because
(i) it would require a remarkable coincidence for the low-ions and
high-ions to arise in truly distinct `clouds' yet share the same
velocity;
and
(ii) the low \nhi\ of this LLS requires high ionization fractions
and corresponding \ion{Si}{4}, \ion{C}{4} absorption unless one
invokes a very large gas density $n_H \gtrsim 0.1 \cm{-3}$
(see Appendix~\ref{sec:2phase}).
We attribute the differences in $b$-values to the effects of photoionization 
from external sources onto an optically thick, self-shielding gas.  
Optical depth effects imply that the lower ionization states
lie preferentially toward the middle of the absorbing gas while the
high-ion gas arises in the outer regions.  
In this scenario, one expects a temperature
gradient which would imply higher $b$-values for the high-ions. 
One can also allow for (and may even expect)
a gradient in velocity dispersion that
would further boost the Doppler parameters.
A full treatment of such a model is beyond the scope of this paper.
We proceed assuming a single-phase (constant density), optically thick model
and restrict the ionic ratios to the gas at the velocities 
$\delta v = -162, -30, +126 \mkms$ as listed in Table~\ref{tab:llsvpfit}.
In Appendix~\ref{sec:2phase}, we explore
how the results vary if we assume that the majority
of the high-ion absorption is
unrelated to the low-ion gas.
We find that the principal scientific conclusions are qualitatively the same.

\begin{figure}
\begin{center}
\includegraphics[width=3.5in]{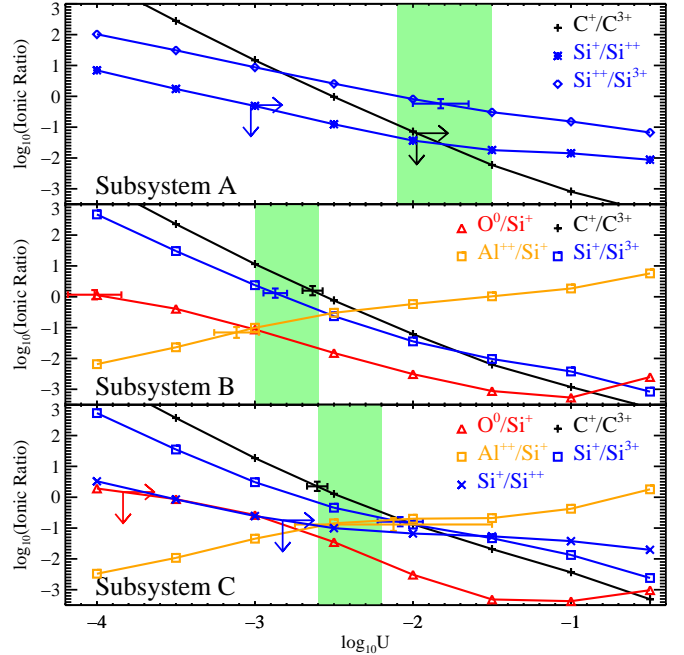}
\end{center}
\caption{
These curves show the predictions from a series of Cloudy photoionization
models for the column density ratios of Si, C and O atoms and ions.
The models are parameterized by the ionization parameter $U$ which corresponds
to the ratio of ionizing photons to hydrogen atoms per unit volume.
In each panel, we have tuned the Cloudy models to match the observed
\nhi\ values and metallicities of subsystems A, B, and C from the LLS at
$z \approx 3.55$.
Overplotted on the curves are the observational constraints.  We
have used these to estimate the ionization parameter for each subsystem
(designated by the shaded regions in each panel). 
  \label{fig:llscldy}}
\end{figure}

In Figure~\ref{fig:llscldy} we present the predicted ratios for
pairs of ions observed in the LLS.  
Overplotted on the curves are the observed values
(assuming a minimum uncertainty of 0.15\,dex) where the
horizontal error bars indicate the
implied constraints for the ionization parameter.
Allowing for the significant uncertainties inherent to this modeling
(simplistic geometry, uncertain atomic data, etc.),
the observations are broadly consistent by taking 
$\log U_A = -1.8 \pm 0.3, \log U_B = -2.8 \pm 0.2, $ and 
$\log U_C = -2.4 \pm 0.2$ for the three subsystems.  
These values are typical of the ionization parameters derived for
a handful other LLSs to date \citep[e.g.][]{bt_1937,p99,pb99,QPQ3}.
These estimated $1\sigma$ uncertainties are dominated 
by systematic (not statistical) uncertainty.
Note that the results do not change even if we assume that
90\%\ of the high-ions arise in an alternate phase (Appendix~\ref{sec:2phase}).
This is because the curves involving \ion{Si}{4} and \ion{C}{4} are 
steep functions of $\log U$.  We also note that the Al$^+$/Si$^+$ 
observations indicate a significantly ionized gas even for our assumed
solar intrinsic Al/Si abundance.

All of the observed ionic ratios are roughly consistent with the assumed
$U$ values save one: 
the O$^0$/Si$^+$ ratio for subsystem~B.  In this subsystem, the observed
column density ratio is $\log[\N{O^0}/\N{Si^+}] = +0.03 \pm 0.1$
yet the photoionization model
predicts $\N{O^0} \ll \N{Si^+}$ for solar relative abundances
and $\log U_B > -4$.  At our preferred value of $\log U_B = -2.80$,
the disparity between model and observation exceeds {\it one order of magnitude}.
We have carefully considered the possibility that the absorption line
at $\lambda = 5925$\AA\ identified as \ion{O}{1}~1302 is a
mis-identification, especially in light of its non-detection 
in subsystem~C.  The line, however, 
is not associated with the other subsystems, the SLLSs at $z \sim 3.19$,
the \ion{Mg}{2} absorber at $z \approx 2.03$, \ion{C}{4}~1548, nor
any other frequently observed doublet of the IGM.  We are 
confident in the identification, therefore, but are challenged to 
explain the offset between observation and model in the O$^0$/Si$^+$ ratio.  
One possibility
is that the gas has a highly super-solar O/Si ratio, i.e., [O/Si]~$\gg 0$.
This runs contrary,  however, to theoretical expectation (and
empirical observations) that $\alpha$-elements 
roughly trace one another because they are both produced
mainly by massive stars. At most, one might allow for departures from the
solar abundances of $\approx 0.3$\,dex.  

\begin{deluxetable}{lccc}
\tablewidth{0pc}
\tablecaption{ELEMENTAL ABUNDANCES FOR THE LLS AT $z \approx 3.55$\label{tab:xh355}}
\tabletypesize{\footnotesize}
\tablehead{\colhead{Ion} &
\colhead{[X/H]} & \colhead{[X/Si$^+$]}}
\startdata
\cutinhead{Subsystem A$^a$}
C$^{+}$ & $<+ 0.08$ & $<-0.91$ \\
C$^{+3}$ & $-0.33\pm0.50$ & $-1.32\pm0.63$ \\
O$^{0}$ & $<+ 4.00$ & $<+ 3.01$ \\
Al$^{++}$ & $<+ 0.53$ & $<-0.47$ \\
Si$^{+}$ & $<+ 0.99$ & $$ \\
Si$^{+3}$ & $-0.16\pm0.24$ & $-1.15\pm0.37$ \\
Fe$^{+}$ & $<+ 3.86$ & $<+ 2.87$ \\
\cutinhead{Subsystem B$^b$}
C$^{+}$ & $-0.66\pm0.20$ & $-0.05\pm0.20$ \\
C$^{+3}$ & $-0.10\pm0.48$ & $+ 0.50\pm0.49$ \\
O$^{0}$ & $>-0.14$ & $>-1.08$ \\
Al$^{++}$ & $-0.95\pm0.20$ & $-0.35\pm0.20$ \\
Si$^{+}$ & $-0.61\pm0.20$ & $$ \\
Si$^{+3}$ & $-0.73\pm0.40$ & $-0.13\pm0.40$ \\
Fe$^{+}$ & $<+ 0.45$ & $<+ 1.05$ \\
\cutinhead{Subsystem C$^c$}
C$^{+}$ & $-1.89\pm0.20$ & $+ 0.30\pm0.20$ \\
C$^{+3}$ & $-2.34\pm0.43$ & $-0.14\pm0.35$ \\
O$^{0}$ & $<-0.36$ & $<+ 1.83$ \\
Al$^{++}$ & $-2.26\pm0.20$ & $-0.07\pm0.20$ \\
Si$^{+}$ & $-2.19\pm0.20$ & $$ \\
Si$^{+3}$ & $-1.85\pm0.33$ & $+ 0.34\pm0.25$ \\
Fe$^{+}$ & $<-0.65$ & $<+ 1.54$ \\
\enddata
\tablenotetext{a}{Assumes a Cloudy photoionization model with $\mnhi = 10^{16.7} \cm{-2}$, [M/H] = -0.50 and log $U = -1.8\pm 0.3$\,dex}
\tablenotetext{b}{Assumes a Cloudy photoionization model with $\mnhi = 10^{16.9} \cm{-2}$, [M/H] = -0.50 and log $U = -2.8\pm 0.2$\,dex}
\tablenotetext{c}{Assumes a Cloudy photoionization model with $\mnhi = 10^{17.9} \cm{-2}$, [M/H] = -1.50 and log $U = -2.4\pm 0.2$\,dex}
\tablecomments{In all cases, we have assumed a minimum error of 0.2\,dex
due to systematic errors in the photoionization modeling.}
\end{deluxetable}

The predicted low value for the O$^0$/Si$^+$ ratio from our models
with $\log U > -4$ stem primarily from the large cross-section
of O$^0$ to extreme ultraviolet and x-ray photons \citep{sj98}.
If one adopts a softer spectrum (i.e.\ one absent the influence of
quasars), the O$^0$/Si$^+$ ratio tends toward the intrinsic O/Si
abundance, e.g. $\log[\N{O^0}/\N{Si^+}] \approx +1$ for solar relative
abundances.  It is possible that the observed ratio,
$\log[\N{O^0}/\N{Si^+}] \approx 0$, indicates a softer radiation field
than we have adopted.  
This would require an intense and local source of radiation, e.g.\ 
the UV flux from a star-burst galaxy.
Deep imaging of the field does not reveal any nearby bright source
of UV flux \citep{ock06}.   We proceed by adopting
the `quasar+galaxy' EUVB model and report the O/H abundance
as a lower limit from the observed O$^0$/H$^0$ ratio.  
Interestingly, this yields [O/H]~$>-0.1$\,dex, a limit that
lies three times above the ionization corrected [Si/H] value.
We note that a softer ionizing spectrum would also imply a higher 
Si/H abundance for subsystem~B. 
Table~\ref{tab:xh355} presents the absolute and 
relative abundances for the LLS adopting the ionization
corrections derived from the `quasar-galaxy' EUVB photoionization 
model with the exception of oxygen in subsystem~B where we adopt conservative
lower limits based on the observed O$^0$/H$^0$ ratio.  
We discuss these results at further
length in the following section.

Under the assumption of detailed balance, we can set an 
upper limit to the electron density of the gas from the
non-detection of \ion{C}{2}*~1335 absorption in subsystems
B and C.  Adopting an electron temperature $T_e = 25,000$K,
which is appropriate for this photoionized gas,
we have $n_e = 117 / [2 \N{C^+_{J=1/2}}/\N{C^+_{J=3/2}} - 1]$.
The upper limits to $\N{C^+_{J=3/2}}$ from the non-detections
of the \ion{C}{2}*~1335 transition imply $n_e < 13 \cm{-3}$
and $<5 \cm{-3}$ for subsystems B and C respectively\footnote{
Note that LLSs with larger $\rm C^+$ column densities should
provide tighter (more meaningful) constraints on $n_e$.}.
Because the gas is predominantly ionized, we infer the same
upper limits to the hydrogen volume densities $n_H$.
These results are independent of the assumed ionization
model.
Finally, we can estimate lower limits to the characteristic
sizes $\ell \equiv N_H / n_H$ of the `clouds' comprising 
subsystems B and C: $\ell_B > 0.5$\,pc and $\ell_C >28$\,pc, respectively.

\begin{deluxetable}{cccc}
\tablewidth{0pc}
\tablecaption{SUMMARY OF PROPERTIES FOR THE LLS AT $z \approx 3.55$\label{tab:llssumm}}
\tabletypesize{\footnotesize}
\tablehead{\colhead{Property} &\colhead{A} & \colhead{B} & \colhead{C} }
\startdata
log (\nhi/$\cm{-2}$) &$14.9\pm0.20$&$16.82\pm0.20$&$18.0\pm0.25$\\
log U &$-1.8\pm 0.3$&$-2.8\pm 0.2$&$-2.4\pm 0.2$\\
$\log (1-x)^a$ &$-3.42_{-0.35}^{+0.36}$&$-2.46_{-0.21}^{+0.21}$&$-2.73_{-0.22}^{+0.23}$\\
$\log (N_{\rm H}/\cm{-2})$ &$18.30\pm0.41$&$19.28\pm0.29$&$20.73\pm0.34$\\
$n_e \; (\cm{-3})$ & &$<13.0$&$< 5.0$\\
$n_H \; (\cm{-3})$ & &$<13.0$&$< 5.0$\\
$\ell \; ({\rm pc})$ &&$>  0.5$&$> 34.8$\\
$\lbrack$O/H]&$<  1.46$&$> -0.2$&$< -1.65$\\
$\lbrack$Si/H]&$ -0.10 \pm 0.35$&$ -0.61 \pm 0.20$&$ -2.19 \pm 0.29$\\
$\lbrack$C/H]&$ -0.17 \pm 0.60$&$ -0.66 \pm 0.20$&$ -1.89 \pm 0.25$\\
$\lbrack$Fe/H]&$<  3.78$&$<  0.45$&$< -0.65$\\
\enddata
\tablenotetext{a}{The ionization fraction $x$ is defined as H$^+$/H.}
\tablecomments{Chemical abundances [X/H] assume the photoionization models as described in Table~\ref{tab:xh355} except for
O/H which adopts no correction.}
\end{deluxetable}

\section{Discussion}

In the previous sections, we presented measurements of the ionic
column densities of the Lyman limit system at 
$z \approx 3.55$ along the sightline to \qso.
We then compared these values against photoionization models
of plane-parallel gas slabs to infer the ionization state of the
gas and thereby estimate physical conditions and chemical abundances
for the subsystems.  A summary of the key properties is
presented in Table~\ref{tab:llssumm}, under the assumption of
a simplistic, single-phase photoionization model.
Appendix~\ref{sec:2phase} discusses how these results change if
we adopt more complex models.

A principal result of this work is the precise measurement of the
\ion{H}{1} column density of a LLS with $\mnhi \approx 10^{18} \cm{-2}$
through a combined analysis of the Lyman limit and the full set
of Lyman series lines.
We have demonstrated that echelle spectra of the full Lyman series 
and the Lyman limit can constrain \nhi\ in the dominant 
low-ion components (assuming identical redshifts)
to several tenths dex for an absorber which lies on the 
flat portion of the curve-of-growth and which is significantly blended
with the $z \sim 3$ IGM and several foreground LLSs.
Presently, there is very weak empirical constraint on the frequency distribution
of \nhi\ values \fnhi\ for the LLS and $\mnhi < 10^{19} \cm{-2}$.
The results presented here give confidence that one can establish
\fnhi\ for the LLS provided a large survey of systems with echelle
observations using the full Lyman series.  This forms the basis
of an observational survey we have recently completed.

More importantly,
we have resolved the LLS system into several
metal-bearing velocity intervals  (termed subsystems A, B, and C) 
and independently constrained their \nhi\ values, ionization state,
and physical properties.  This enables a
probe of metallicity variations at $\delta v \lesssim 100 \mkms$ 
velocity separation
and in gas associated with a single galaxy halo or galaxy-scale 
structure\footnote{The association of a Lyman limit system
with a single galactic structure is well-motivated
by the rarity of LLSs along individual sightlines.}.
As Table~\ref{tab:llssumm} reveals, these subsystems exhibit large differences
in enrichment ($\gtrsim 1$\,dex).  Although we may dismiss subsystem~A
from the discussion because of its much lower \nhi\ value, the metallicities
of subsystems~B and C alone diverge by over an order of magnitude.
Although large dispersions in gas metallicities have been reported for
the intergalactic medium \citep[e.g.][]{schaye03} and individual
gas-rich galaxies at high $z$ \citep{pgw+03,pdd+07}, 
these studies refer to gas in structures separated by very
large distances ($\gg 1$\,Mpc).  Our results,
which mark the first detailed analysis of 
multiple components in a LLS drawn from a homogeneous 
sample\footnote{There are also reports of metallicity
variations in the few special LLS that permit D/H analysis \citep{tfb96,kts+03}.
\cite{pb99} also presented similar results for two
subsystems in the partial LLS at $z=1.92$ toward J2233$-$606 noting
a difference of at least 0.5\,dex in [C/H].}, 
indicate a similar dispersion holds down to galactic scales.
This conclusion resembles that of \cite{sck07}, who suggest
the IGM is inhomogeneously enriched by a population
of high metallicity, \ion{H}{1}-deficient absorbers\footnote{One may note
that subsystem~A could be a higher density analog of these 
\ion{H}{1} deficient metal-line systems.}.
The large metallicity variations are also consistent with the small
physical sizes implied for metal-line systems from sightline
studies of lensed quasars \citep{rsb+01}.

\begin{figure}
\begin{center}
\includegraphics[height=3.5in,angle=90]{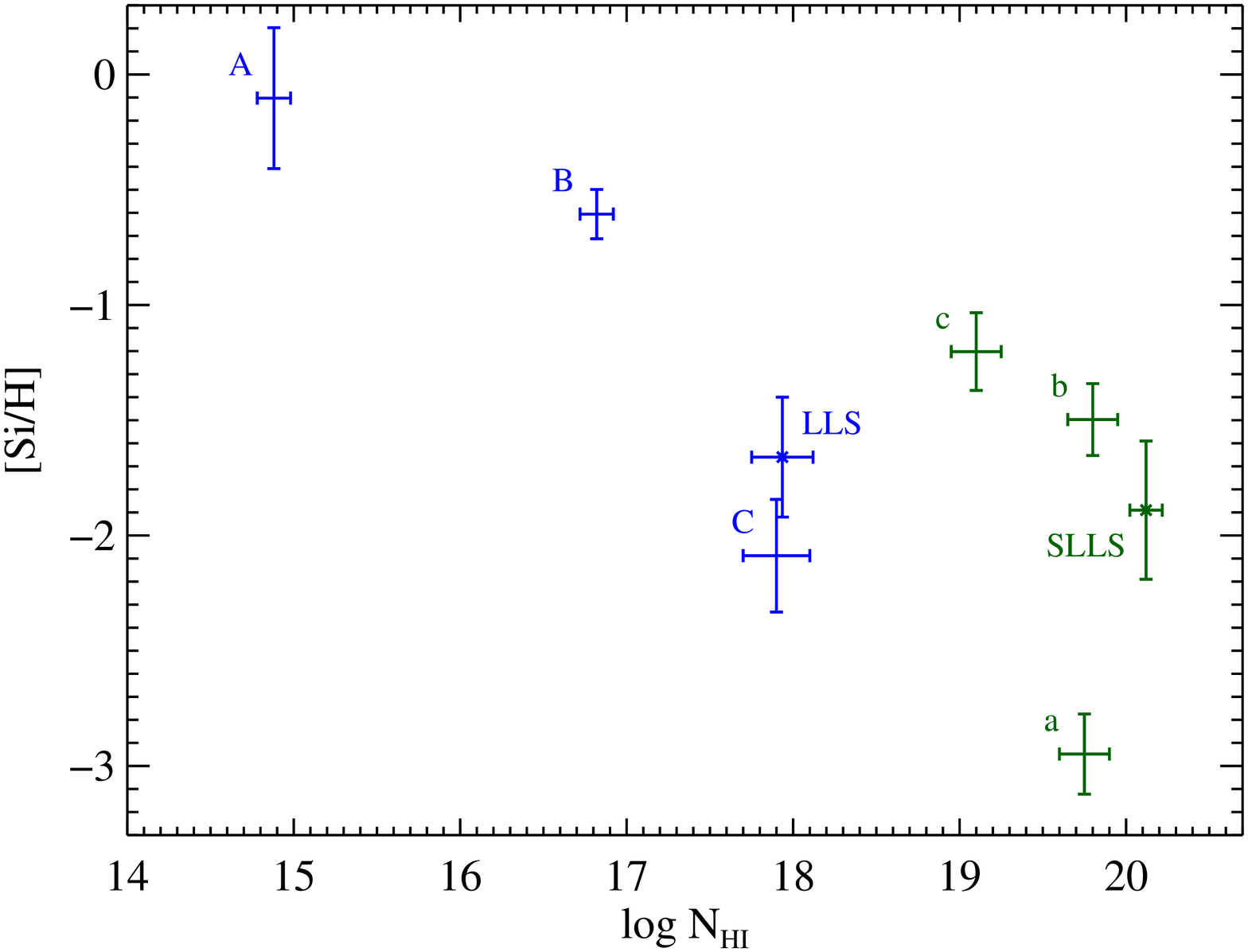}
\end{center}
\caption{Silicon metallicity [Si/H] versus \ion{H}{1} column density 
\nhi\ as measured for the three subsystems (A,B,C) of the LLS at
$z \approx 3.55$ and the three SLLSs (a,b,c) at $z \approx 3.19$.
For each system, the figure also presents the $\mnh$-weighted, mean
metallicity $\rm <[Si/H]>$
as an additional error interval marked with an `x'.
There is an apparent decrease in [Si/H] with increasing \nhi\ value.
This is partly due to observational bias: metal-poor systems with low
\nhi\ value were not considered in this paper.  
The large dispersion in metallicity between LLSs and also in 
the velocity components comprising
a given LLS is a robust result of our analysis.
  \label{fig:mtlvsNHI}}
\end{figure}

We have also analyzed the metallicity of the SLLSs at $z \approx 3.19$
toward \qso\ (Appendix~\ref{sec:appx}, Table~\ref{tab:sllssumm}).
These are separated by several hundred \kms\ and are more likely 
to correspond to multiple galactic halos and/or large-scale
structures.  
The observational results for the LLS, its subsystems, and the SLLSs
are presented in 
Figure~\ref{fig:mtlvsNHI}
where the metallicity values are plotted against the measured 
\ion{H}{1} column densities.
At face value, the figure reveals a trend of declining metallicity with
increasing \nhi\ value.   Although this is an accurate account of the
systems analyzed in our paper, the trend is biased
by our having focused on metal-bearing gas.  
At the lower \nhi\ values, especially,
there are \ion{H}{1} components in the LLS (and presumably the SLLSs)
that exhibit only weak
or non-discernible metal-line absorption.  An obvious example of
this is the additional components within subsystem~B of the LLS
(Table~\ref{tab:llsvpfit}).
These have estimated \nhi\ values of nearly $10^{17} \cm{-2}$ yet 
show essentially no low-ion absorption and only weak high-ions
(Figure~\ref{fig:z355mtl}).  We have not estimated metallicities
for these components, in part because we lack constraints
on the ionization state. 
Nevertheless, this gas likely has significantly 
lower metallicity than subsystem~B and possibly even subsystem~C.
It is our expectation, therefore, that the low \nhi, low metallicity
region of Figure~\ref{fig:mtlvsNHI} is populated by gas related to the LLS.
The most robust result regarding the metallicities, therefore,
is that the gas in LLSs exhibits a large dispersion (i.e.\ greater
than 1\,dex). 
This holds not only for the 
subsystems comprising the LLS but also among the various LLSs
studied here. 
%Under the assumption that these subsystems correspond to the same
%galactic or galaxy-scale structure (an assumption well-motivated
%by the rarity of LLSs along individual sightlines),
%this implies that metals are not well-mixed on the scales
%probed by LLS absorption, i.e.\ tens of kpc.  

The most natural interpretation of the large metallicity
variations is the incomplete mixing of metals in the associated
galactic-scale structure.
Presumably the gas has previously cycled through the neutral ISM
of a nearby galaxy\footnote{An inspection of deep optical
images of the field surrounding \qso\ does not reveal any 
obvious $z \sim 3.5$ galactic counterpart \citep{ock06}.}
(i.e.\ a damped \lya\ system) and subsequently
was transported to the surrounding region.  One plausible
explanation is that this transport process does not efficiently mix
the gas. 
Another possibility is that the large metallicity variations represent
a homogeneously enriched material that has mixed
inhomogeneously with `fresh'
primordial gas.  In both scenarios, we conclude that mixing is incomplete
on galactic scales and that large metallicity variations may occur
in the galaxy/IGM interface.

The large metallicity variation observed within
the LLS may also have important implications for larger 
\nhi\ systems where the \ion{H}{1} absorption is not resolved
and, therefore, metallicity gradients cannot be assessed.
Metallicity variations within the 
SLLSs and DLAs could occur if the sightlines sample a
metallicity gradient within the galaxy and/or multiple gas phases
with varying enrichment levels surrounding high $z$ galaxies
\citep{wp00a,fpl+07,lhp+08}.
A large metallicity dispersion within DLAs
runs contrary, however, to the relatively
small dispersion observed in the relative abundances of
$\alpha$/Fe, N/O, etc.\ \citep{pro03,dz06}.  
Nevertheless,
we are motivated to explore metallicity variations in a larger sample
of LLSs and to seek trends with the total \nhi\ value
of the system.  
This is of special interest given that 
the $N_{\rm H}$ value of the LLS
and SLLS is representative of DLAs.  The principal physical
difference could simply be the ionization parameter of the gas.

Figure~\ref{fig:mtlvsNHI} also presents the $N_{\rm H}$-weighted,
mean metallicity integrated over the entire LLS and SLLS systems, 
i.e. 

\begin{equation}
<{\rm [Si/H]}> = \frac{\smm_i {\rm [Si/H]^i} N_{\rm H}^i}{\smm_i N_{\rm H}^i} 
\end{equation}
where $N_{\rm H}^i$ was derived from photoionization modeling.
Regarding the LLS at $z \approx 3.55$, we find that its mean
metallicity is dominated by subsystem~C which has the lowest
level of enrichment but the highest $\mnh$ value\footnote{We
expect that the inclusion of other \ion{H}{1} clouds related to the
LLS, e.g.\ the metal-poor components in subsystem~B, would not
significantly modify this result for they will have significantly 
lower $\mnh$ value.}.
If we had derived ionization corrections and abundances
for the integrated \ion{H}{1} and
ionic column densities of the LLS, we would have derived a
lower ionization potential, a lower total $\mnh$ value ($\approx 0.3$\,dex), 
and a correspondingly higher metallicity.  
If metal-bearing clouds consistently exhibit lower metallicity in
higher \nhi\ components (Figure~\ref{fig:mtlvsNHI}), then 
one may systematically overestimate the metallicities
of the LLS by analyzing integrated column
densities.  It is also possible that such a systematic effect afflicts current
studies of the predominantly ionized SLLSs \citep{pdd+07,poh+06}. 
In a predominantly ionized gas,
it is of primary importance
to precisely constrain the ionization state of the component
that dominates the \ion{H}{1} column density (here subsystem~C)
because this gas may dominate the total $\mnh$ value.

Another important result presented here relates to using
the \ohr\ ratio to assess the metallicity of LLSs.
It has been conventional wisdom in the ISM and QAL communities
that the \ohr\ ratio is an excellent predictor of O/H abundance because: 
(i) charge exchange reactions between O$^0$ and H$^0$ imply
similar ionization fractions for these atoms 
(i.e.\ O$^0$/O $\approx$ H$^0$/H); and
(ii) oxygen contributes significantly to the metal
mass density of chemically enriched gas.
The first point is valid even for gas subjected
to an intense ionizing radiation field
but only for radiation with a soft spectrum (stellar dominated).
A hard radiation field (e.g.\
quasar dominated) is predicted to overionize O$^0$ with respect to 
H$^0$ because O$^0$ has a higher cross-section to hard photons
\citep{sj98}.  As Figure~\ref{fig:llscldy} indicates,
a highly ionized gas subjected to a hard radiation
field (e.g.\ an LLS subjected to the EUVB) should exhibit an
\ohr\ ratio that underpredicts
the true O/H value.  This behavior is evidenced in subsystem~C
of the LLS where we report the positive detection of \ion{Si}{2}~1304
while no absorption from \ion{O}{1}~1302 is measured.
In this case, the \ohr\ ratio may underpredict the O/H abundance.
In subsystem~B, however, we tentatively
report the positive detection of \ion{O}{1}~1302
which suggests [O/Si]~$>+1$\,dex if we assume a quasar+galaxy
radiation field.  
This large O/Si ratio suggests the gas is photoionized by 
a softer ionizing spectrum, which would require an intense,
local starburst.  We note again, however, that we cannot identify
any such source in our images of this field \citep{ock06}.  
In future papers, we will explore further the validity
of the \ohr\ ratio as a metallicity estimator.

%[Discuss Figure~\ref{fig:nmtlvsnh}]

Before concluding, we place the metallicities of the LLSs 
toward \qso\ within the context of the cosmic metal budget
of the $z \approx 3$ universe.  The $\mnh$-weighted mean metallicities
of the LLS at $z \approx 3.55$ and the SLLSs at $z \approx 3.19$
are $\rm <[Si/H]> = -1.66 \pm 0.25$ and
$\rm <[Si/H]> = -1.64 \pm 0.30$ respectively.
These central values are in good agreement with the mean metallicity
of DLAs at $z \approx 3$ \citep{pgw+03}, i.e.\ the neutral,
atomic ISM gas of high $z$ galaxies.
The total $\mnh$ of our systems, meanwhile, is comparable to the
median \nhi\ value of the DLAs \citep{phw05}.  This implies
that each of the LLS contributes as many metals as a typical DLA
system.  The key difference, of course, is that the LLSs are
roughly $12\times$ more common than DLAs (Paper~I).
If the LLSs studied here are characteristic of the full LLS
population (a wild speculation), the LLSs would contain $\approx 12\times$
the mass in metals as the DLAs.
In this case, the LLSs would represent the dominant reservoir of metals
in the universe.  Taking this speculation to the limit, we 
conclude that the majority of metals in our universe
lie at the galaxy/IGM interface.  This is analogous to the notion
that the metals in galaxy clusters resides primarily in the intracluster
medium.

\section{Summary}

In this manuscript, we have analyzed the \nhi\ value, ionization
state, and chemical abundances of the subsystems comprising
a Lyman limit system (LLS) at $z \approx 3.55$ toward \qso\
(Table~\ref{tab:llssumm}).
We demonstrate that the \nhi\ value of an absorber whose 
\lya, \lyb, and \lyg\ lines lie on the flat portion of the curve-of-growth
can be constrained to within a few tenths dex
provided analysis of the Lyman limit and the full Lyman
series when resolved by echelle observations.
As important, one can resolve the LLS into subsystems with distinct
kinematics and perform photoionization and abundance analyses for gas on
(presumable) scales of several to tens kpc.
The subsystems exhibit a presumed significant metallicity variation
($>1$\,dex) indicating incomplete mixing of metals in the gas
comprising a galactic-scale structure.
Finally, the following Appendix provides measurements of the
physical properties for the SLLSs at $z\approx 3.19$ along
the quasar sightline.  Future papers will present the \nhi\ and
abundance measurements for a large sample of $z \sim 3$ LLSs.

\acknowledgments

This paper includes data gathered with the 6.5 meter Magellan Telescopes 
located at Las Campanas Observatory, Chile.
GEP and JXP were supported by NSF grant AST-0307408.  
JO and SB acknowledge support from NSF grant AST-0307705.

%%%%%%%%%%%%%%%%%%%%%%%%%%%%%%%%%%%%%%%%%%%%%%%%%%%%%%%%%%%%%%%%%%%%%%%%
\appendix

\section{A Two-Phase Scenario for the LLS at $z \approx 3.55$}
\label{sec:2phase}

The results presented in the paper on the LLS at $z \approx 3.55$ 
were derived using a simplistic, single-phase photoionization model.
This model assumes all of the gas is co-spatial, has constant
density, has similar temperature, and experiences the same radiation field.
The data, however, offer some indications for a more complex, 
multi-phase medium.  First, the profile fits presented reveal systematically
larger $b$-values for the high-ion gas than the low-ion gas.  This 
difference may indicate that some of the high-ion gas occurs in a 
separate phase from the low-ion gas, even though they have the same
central velocity.  We caution, however, that we can construct a model
where the high-ion gas which coincides in velocity with the low-ion
gas has identical $b$-value.  The other observation that suggests
a multi-phase medium is the positive detection of \ion{O}{1} absorption
in subsystem~B and the correspondingly large observed O$^0$/Si$^+$ ratio.
As described in the text, this is difficult to accommodate within a
single-phase model if we assume the intrinsic O/Si ratio has a roughly
solar relative abundance.

\begin{figure}
\begin{center}
\includegraphics[height=6.8in,angle=90]{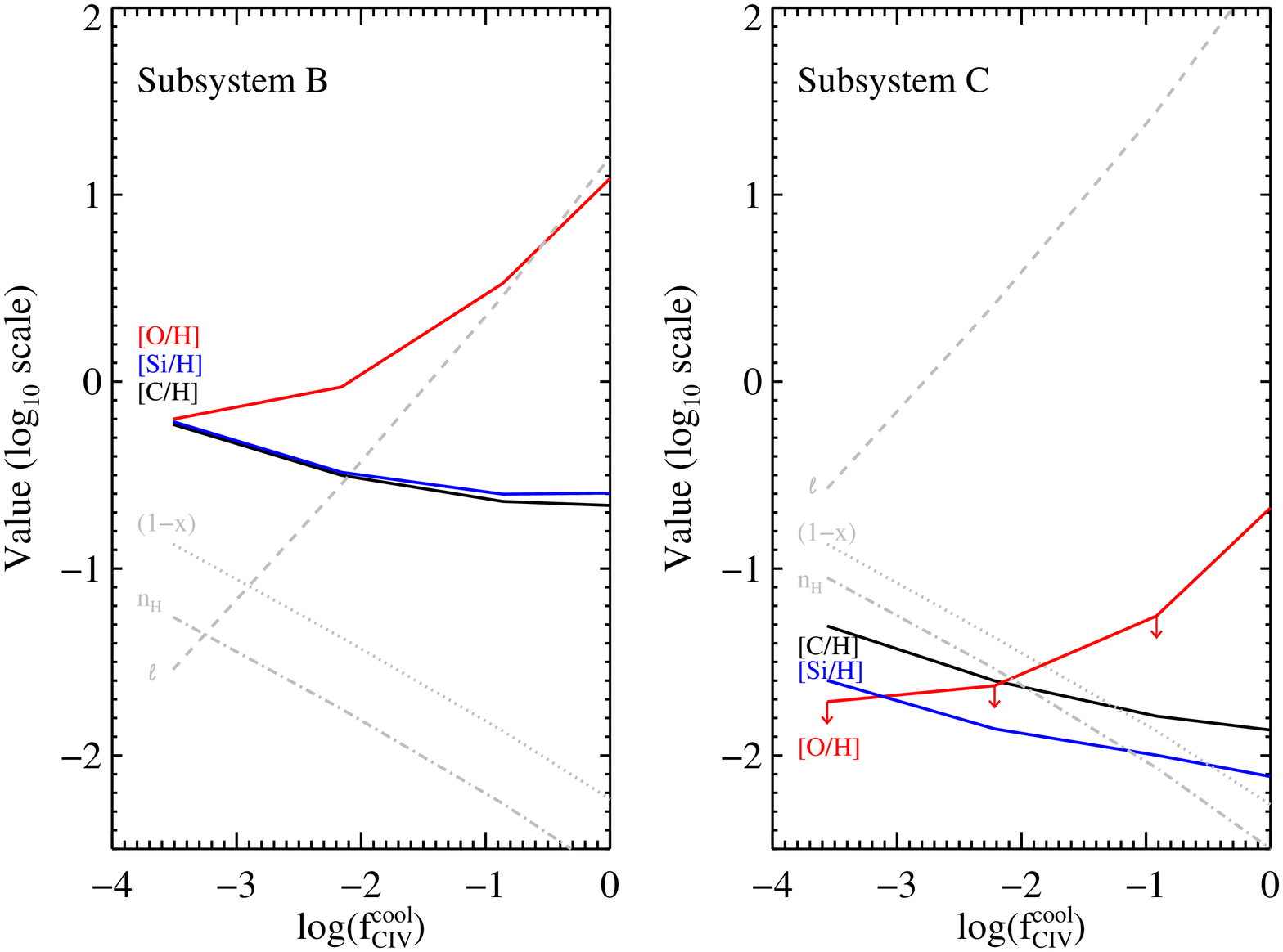}
\end{center}
\caption{Modeled results for the abundances (solid curves) and 
physical conditions (non-solid) for the gas in the cooler phase of
a 2-phase medium; see the text for details.  
For this cool phase, we assume photoionization 
equilibrium and that this phase contributes the fraction of 
\ion{C}{4} column density indicated on the x-axis.
The results for $\log(\mfciv) = 0$ correspond
to the single-phase scenario described earlier in the text 
($\S$~\ref{sec:photo}).  For the scenarios where the cool phase
contributes less than $1\%$ of the observed \ion{C}{4}, the
density of the cool phase must exceed $n_{\rm H} > 0.01\,\cm{-3}$
giving clump sizes less than 10\,pc.  For Subsystem~C, we would
predict anomalously low O/C ratios.
\label{fig:2phase}}
\end{figure}

With these motivations in mind, we performed the following calculation
for subsystems~B and C.   We adopted a two-phase model (cool and warm)
where the cool phase is presumed to be a photoionized gas
that gives rise to all of the observed low-ions but only a fraction
\fciv\ of the high-ion gas.  The warm phase contributes the remainder
of the high-ions observed and is otherwise ignored in this analysis.
In Figure~\ref{fig:2phase}, we summarize the implications of this
two-phase model as a function of \fciv.  The results at $\log \mfciv = 0$
correspond, by definition, to the single-phase models considered in 
$\S$~\ref{sec:photo}.  

The solid curves in Figure~\ref{fig:2phase} show the chemical abundances
of O, C, and Si in the cool phase.  We note a strong dependence of O/H
on \fciv\ which emphasizes the fact that the observed O$^0$/H$^0$ ratio
is not a robust measure of O/H in highly ionized systems.
In contrast, the Si/H and C/H abundances (which are derived from the 
Si$^+$/H$^0$ and C$^+$/H$^0$ ratios) have only mild dependence on \fciv.
At the lowest \fciv\ values considered, which correspond to $\log U \approx -4$
in the cool phase, we observe for subsystem~B that the O, Si, and C abundances
have roughly solar relative and absolute abundances.
The comparison cannot be well made for subsystem~C as the O abundances
are strictly upper limits due to the non-detection of \ion{O}{1}~1302.
We do note, however, that very low values for \fciv\ imply super-solar C/O
ratios. 

The gray dotted, dashed, and dot-dashed curves in Figure~\ref{fig:2phase}
show the physical conditions of the cool phase.  We find that the
neutral fraction $(1-x)$ of the gas remains low ($\lesssim 10\%$)
for the range of \fciv\ values considered.  The densities plotted correspond
to a lower limit on the gas density where we have converted the
ionization potential of the gas into a density by adopting the mean
intensity of the EUVB at $z \approx 3.55$ 
(our input spectrum from CUBA has $\log \Phi_{\rm EUVB} = 5.6$).
If there is an additional, local radiation field then the densities will
be higher than those plotted.  At the lowest \fciv\ values
we find $n_{\rm H} \gtrsim 0.1 \cm{-3}$ which would be a surprisingly
high value for LLS systems although we have little to compare against.
More importantly, the inferred sizes $\ell$ (plotted in units of 100\,pc) 
for $\mfciv < 10{-2}$ are of the order of 10\,pc or smaller.
In contrast, the single-phase solutions imply sizes of the order 1 to 10 \,kpc
for the two subsystems.

If we were to adopt a two-phase scenario for subsystems~B and C where
the majority of high-ions arise in the warm phase, then
we are driven to a rather dense medium with very small physical size.
The odds of intersection such a system would be extremely small unless
there were millions (even billions) of such `clouds' for every high $z$ galaxy.
For this reason (and the anomalous C/O and even Si/O abundances for subsystem~C),
we do not favor such models but these cannot be formally ruled out
by the data.

Finally, we comment on the implications of two-phase scenarios on the
principle conclusions of this paper regarding the LLS system.
Our conclusions on the ionization state (highly ionized) and the
large difference in metal abundances between subsystems~B and C
remain unchanged.  Furthermore, we would still predict a mean metallicity
for the cool gas in the LLS of $\approx 1/30$ solar abundance.
In short, the only major difference between the two-phase and single-phase
scenario is in the inferred size of the absorbing gas.

\section{Analysis of the Super Lyman Limit Systems at $z \approx 3.19$}
\label{sec:appx}

\subsection{Ionic Column Densities}

Strong \lya\ absorption at $\lambda \approx 5080$\AA\ is evident in the
spectrum of \qso\ (Figure~\ref{fig:spec}).  
This absorption was identified and analyzed in Paper~I 
where we modeled the absorption 
as three \lya\ profiles corresponding to three
distinct metal-line complexes at $z \approx 3.19$.  
Each of these \lya\ profiles has \ion{H}{1} column densities
that likely satisfy the \nhi\ criterion of a super Lyman limit
system (SLLS; $\mnhi \ge 10^{19} \cm{-2}$).
The full SLLS criteria introduced in Paper~I, however, 
includes criteria one that one group all \lya\ lines lying within 
$300 \mkms$ of one another into a single SLLS.
We introduced this criterion because fits to \lya\ absorption lines
with velocity separation $\delta v < 300 \mkms$ can be highly degenerate.
In the following, however, we will treat the three fitted
\lya\ lines as separate SLLSs
and refer to them as \aslls, \bslls, and \cslls.
We have done this primarily to investigate metallicity variations
among individual Lyman limit systems but also because the degeneracy
between the two closest \lya\ lines (b,c) is not too severe. 
We caution the reader, however, that \bslls\ and \cslls\ should {\it not}
be considered as separate in any statistical analysis of SLLS, especially
one that draws upon the results of Paper~I.

\begin{figure}
\begin{center}
\includegraphics[height=6.8in]{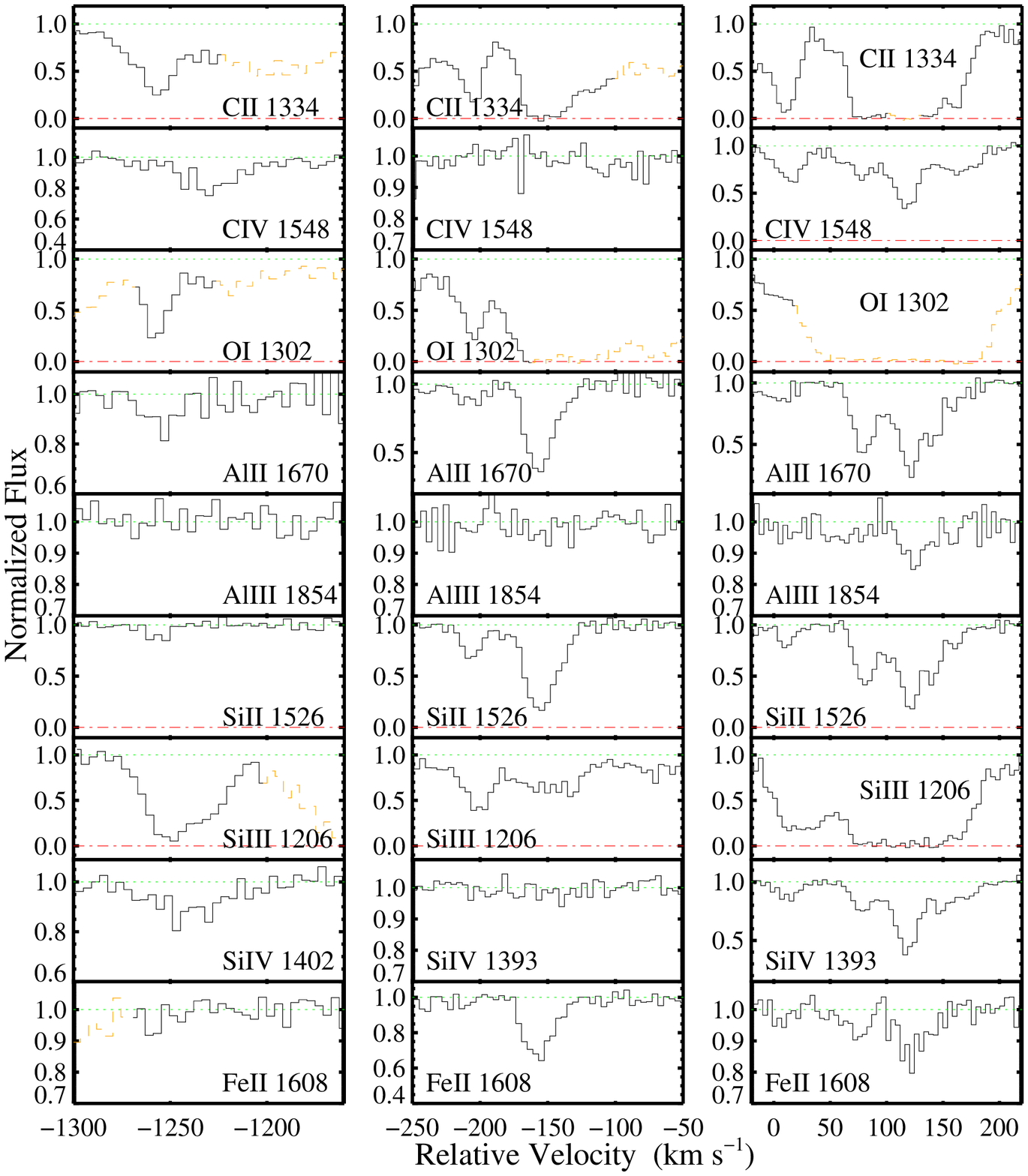}
\end{center}
\caption{Metallic absorption for the SLLSs at $z \approx 3.19$, referred
to in the text as \aslls, \bslls, and \cslls\ (from left to right).
All velocities in the figure are relative to an arbitrary $z=3.190$. 
Blends with coincident absorption (e.g.\ \lya\ lines) are denoted as
(orange) dashed spectra.  
\label{fig:sllsmtl}}
\end{figure}

\begin{figure}
\begin{center}
\includegraphics[height=5.8in,angle=90]{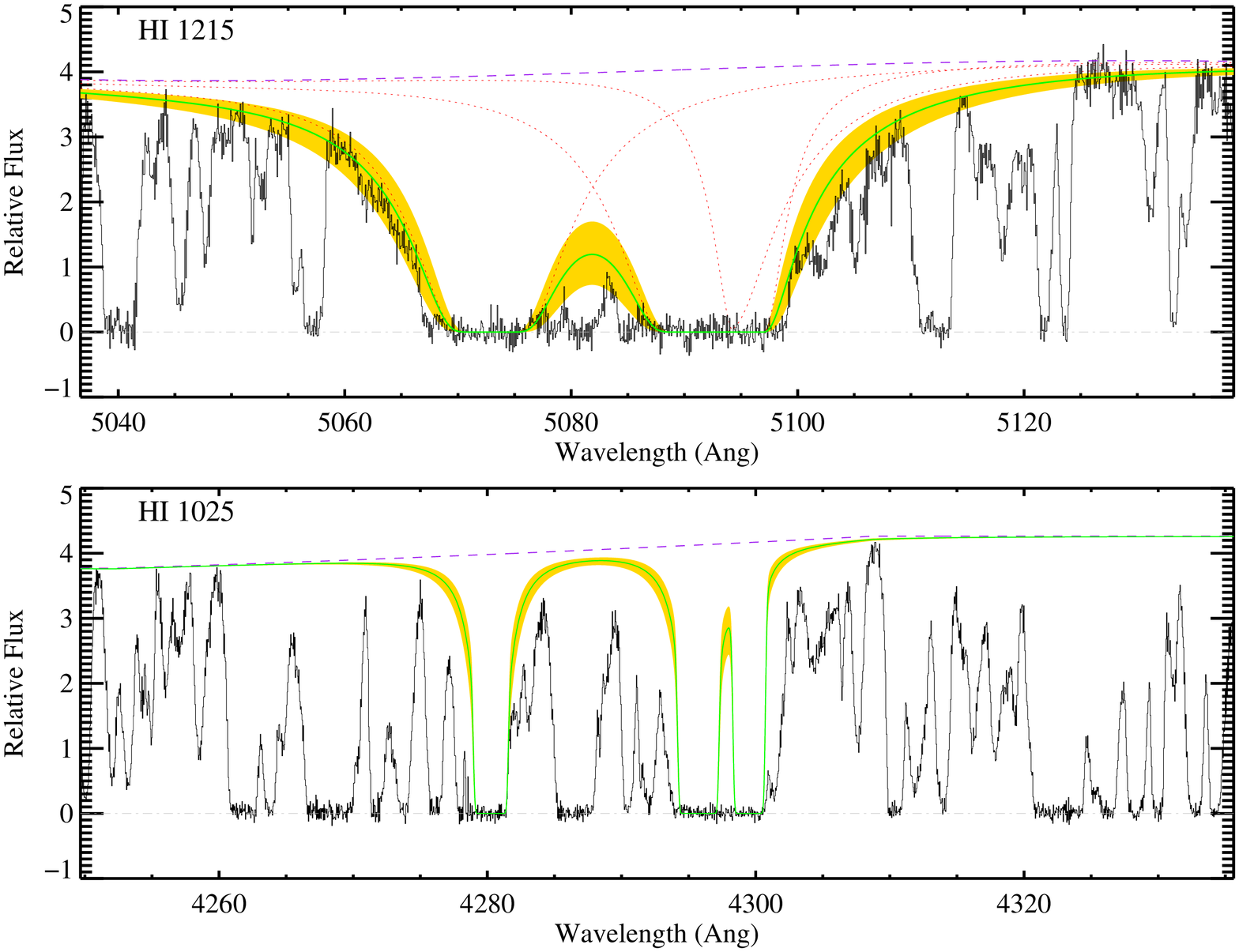}
\end{center}
\caption{Spectral regions covering the \lya\ and \lyb\ profiles
for the SLLSs at $z\approx 3.19$.  The solid (green) overplotted
on the data show the best-fit solution from our line-profile
analysis of these data.  The yellow-shaded region surrounding this
model corresponds the $1\sigma$ uncertainties to the derived
\nhi\ values (Table~\ref{tab:z319clm}).  The (purple) dashed line in each
panel shows our estimate of the local quasar continuum.  Finally,
the (red) dotted lines in the upper panel show the individual \lya\
profiles for \aslls, \bslls, and \cslls.
\label{fig:sllshi}}
\end{figure}

Figure~\ref{fig:sllsmtl} presents a subset of the metal-line 
transitions for the three SLLSs.  We note progressively stronger
low-ion absorption from \aslls\ to \cslls.  
We have measured ionic column densities for these transitions using
the AODM and present the results in Table~\ref{tab:z319clm}.
Following Paper~I (and standard practice), we fit the \lya\ absorption 
with three \lya\ lines centered at the peak optical depth of each
metal-line complex: $z_a = 3.17292, z_b=3.18779, z_c=3.19169$.
Each line exhibits strong damping wings (Figure~\ref{fig:sllshi}),
which provides a relatively precise constraint on the \nhi\ values
(Table~\ref{tab:z319clm}) independently of the assumed Doppler parameters.
Our \nhi\ values are fully consistent with those reported in Paper~I.
Regarding the Doppler parameters, we have only loose constraints from the \lyb\
profiles: $b_a < 35\mkms, b_b < 45\mkms, b_c < 30\mkms$.
The adopted solutions are overplotted in Figure~\ref{fig:sllshi}.
As noted above, there is some degeneracy between the \nhi\ values
of \bslls\ and \cslls.  The \lya\ and \lyb\ data together can be used to 
place an upper limit on the \nhi\ column density of \cslls\ by assuming
there is no significant \lya\ absorption by \bslls: 
$\mnhi^c < 10^{19.3} \cm{-2}$.  Adopting this constraint, we set
a lower limit $\mnhi^b > 10^{19.2} \cm{-2}$ to describe the \lya\
absorption at $\lambda > 5100$\AA.   Furthermore, we derive the
constraint $\log \mnhi^b > \log \mnhi^c - 0.1$.

\begin{deluxetable}{lcccccccc}
\tablewidth{0pc}
\tablecaption{IONIC COLUMN DENSITIES FOR THE SLLS AT $z=3.19$\label{tab:z319clm}}
\tabletypesize{\footnotesize}
\tablehead{\colhead{Ion} & \colhead{$\lambda_{\rm rest}$} & \colhead{$\log f$}
& \colhead{$v_{int}^a$} 
& \colhead{$\log N_{\rm AODM}$}&
  \colhead{$\log N_{\rm adopt}$} \\
& (\AA) & & (\kms) &  }
\startdata
\cutinhead{SLLS$_a$}
\ion{H}{1} & 1215.6701 & $ -0.3805$&&&$19.75\pm 0.15$\\
\ion{C}{2}&1334.5323 &$ -0.8935$&$[-1289,-1234]$&$ 13.88 \pm 0.03$&$ 13.88 \pm 0.03$\\
\ion{C}{4}&1548.1950 &$ -0.7194$&$[-1289,-1229]$&$< 12.81$&$< 12.81$\\
&1550.7700 &$ -1.0213$&$[-1289,-1229]$&$< 13.14$&\\
\ion{O}{1}&1302.1685 &$ -1.3110$&$[-1279,-1239]$&$ 14.15 \pm 0.03$&$ 14.15 \pm 0.03$\\
\ion{Al}{2}&1670.7874 &$  0.2742$&$[-1289,-1229]$&$ 11.59 \pm 0.10$&$ 11.59 \pm 0.10$\\
\ion{Al}{3}&1854.7164 &$ -0.2684$&$[-1289,-1229]$&$< 11.90$&$< 11.90$\\
&1862.7895 &$ -0.5719$&$[-1289,-1229]$&$< 12.24$&\\
\ion{Si}{2}&1190.4158 &$ -0.6017$&$[-1289,-1229]$&$ 12.84 \pm 0.12$&$ 12.89 \pm 0.05$\\
&1193.2897 &$ -0.3018$&$[-1289,-1229]$&$ 12.91 \pm 0.05$&\\
&1304.3702 &$ -1.0269$&$[-1289,-1229]$&$< 13.14$&\\
&1526.7066 &$ -0.8962$&$[-1289,-1229]$&$< 12.65$&\\
\ion{Si}{3}&1206.5000 &$  0.2201$&$[-1299,-1209]$&$ 13.16 \pm 0.03$&$ 13.15 \pm 0.03$\\
\ion{Si}{4}&1402.7700 &$ -0.5817$&$[-1289,-1229]$&$< 12.73$&$< 12.73$\\
\ion{Fe}{2}&1608.4511 &$ -1.2366$&$[-1279,-1229]$&$< 12.86$&$< 12.86$\\
\cutinhead{SLLS$_b$}
\ion{H}{1} & 1215.6701 & $ -0.3805$&&&$19.80\pm 0.15$\\
\ion{C}{2}&1334.5323 &$ -0.8935$&$[ -231, -101]$&$> 14.62$&$> 14.62$\\
\ion{C}{4}&1548.1950 &$ -0.7194$&$[ -251, -101]$&$< 12.59$&$< 12.59$\\
&1550.7700 &$ -1.0213$&$[ -251, -101]$&$< 12.88$&\\
\ion{N}{5}&1242.8040 &$ -1.1066$&$[ -211,  -81]$&$< 13.64$&$< 13.64$\\
\ion{O}{1}&1302.1685 &$ -1.3110$&$[ -251, -201]$&$> 14.26$&$> 14.26$\\
\ion{Al}{2}&1670.7874 &$  0.2742$&$[ -251, -101]$&$ 12.51 \pm 0.03$&$ 12.51 \pm 0.03$\\
\ion{Al}{3}&1854.7164 &$ -0.2684$&$[ -251, -101]$&$< 12.13$&$< 12.13$\\
&1862.7895 &$ -0.5719$&$[ -251, -101]$&$< 12.39$&\\
\ion{Si}{2}&1193.2897 &$ -0.3018$&$[ -251, -101]$&$> 13.83$&$ 13.97 \pm 0.03$\\
&1526.7066 &$ -0.8962$&$[ -251, -101]$&$ 13.97 \pm 0.03$&\\
&1808.0130 &$ -2.6603$&$[ -251, -101]$&$< 14.47$&\\
\ion{Si}{3}&1206.5000 &$  0.2201$&$[ -251, -101]$&$< 13.00$&$< 13.00$\\
\ion{Si}{4}&1393.7550 &$ -0.2774$&$[ -251, -101]$&$< 12.03$&$< 12.03$\\
&1402.7700 &$ -0.5817$&$[ -251, -101]$&$< 12.41$&\\
\ion{Fe}{2}&1608.4511 &$ -1.2366$&$[ -251, -101]$&$ 13.69 \pm 0.04$&$ 13.69 \pm 0.04$\\
\cutinhead{SLLS$_c$}
\ion{H}{1} & 1215.6701 & $ -0.3805$&&&$19.10\pm 0.15$\\
\ion{C}{2}&1334.5323 &$ -0.8935$&$[  -18,  222]$&$> 14.90$&$> 14.90$\\
\ion{C}{4}&1548.1950 &$ -0.7194$&$[  -18,  222]$&$ 13.92 \pm 0.03$&$ 13.92 \pm 0.03$\\
&1550.7700 &$ -1.0213$&$[  -18,  222]$&$ 13.92 \pm 0.03$&\\
\ion{N}{5}&1238.8210 &$ -0.8041$&$[   22,  142]$&$ 13.27 \pm 0.06$&$ 13.27 \pm 0.06$\\
\ion{Al}{2}&1670.7874 &$  0.2742$&$[  -48,  222]$&$ 12.88 \pm 0.03$&$ 12.88 \pm 0.03$\\
\ion{Al}{3}&1854.7164 &$ -0.2684$&$[  -18,  222]$&$ 12.48 \pm 0.08$&$ 12.48 \pm 0.08$\\
&1862.7895 &$ -0.5719$&$[  -18,  222]$&$< 12.49$&\\
\ion{Si}{2}&1190.4158 &$ -0.6017$&$[  -18,  222]$&$ 14.14 \pm 0.03$&$ 14.13 \pm 0.03$\\
&1193.2897 &$ -0.3018$&$[  -18,  222]$&$> 14.15$&\\
&1304.3702 &$ -1.0269$&$[  -18,  262]$&$< 14.36$&\\
&1526.7066 &$ -0.8962$&$[  -18,  222]$&$ 14.13 \pm 0.03$&\\
\ion{Si}{4}&1393.7550 &$ -0.2774$&$[  -18,  222]$&$ 13.30 \pm 0.03$&$ 13.31 \pm 0.03$\\
&1402.7700 &$ -0.5817$&$[  -18,  222]$&$ 13.34 \pm 0.03$&\\
\ion{Fe}{2}&1608.4511 &$ -1.2366$&$[  -18,  222]$&$ 13.49 \pm 0.07$&$ 13.49 \pm 0.07$\\
\enddata
\tablenotetext{a}{Velocity interval for the AODM
relative to $z=3.190000$.}
\end{deluxetable}

\begin{figure}
\begin{center}
\includegraphics[height=5.8in]{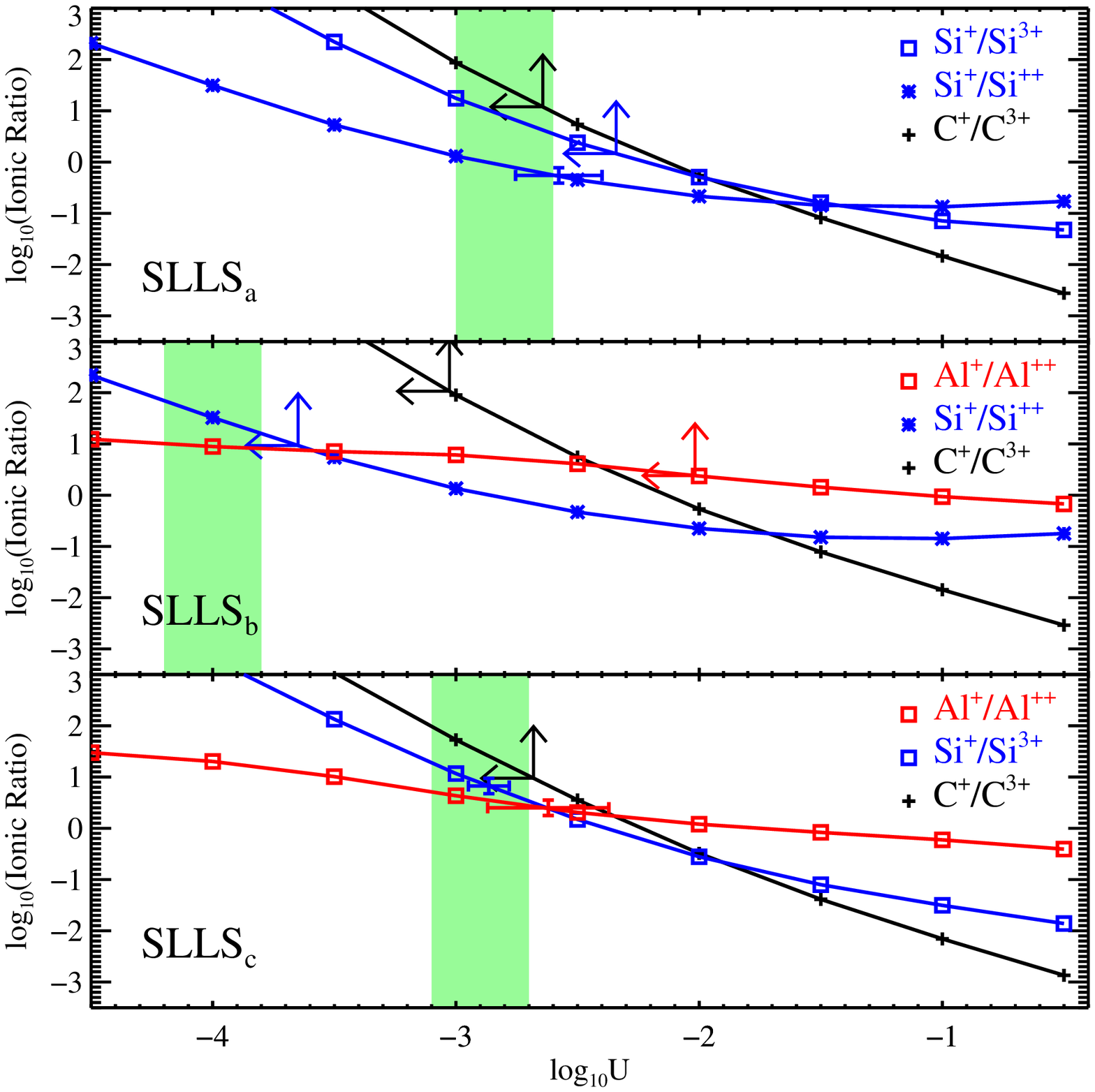}
\end{center}
\caption{
These curves show the predictions from a series of Cloudy photoionization
models for the column density ratios Si, C and Al ions.
The models are parameterized by the ionization parameter $U$ which corresponds
to the ratio of ionizing photons to hydrogen atoms per unit volume.
In each panel, we have tuned the Cloudy models to match the observed
\nhi\ values and metallicities of \aslls, \bslls, and \cslls\ at
$z \approx 3.19$.
Overplotted on the curves are the observational constraints.  We
have used these to estimate the ionization parameter for each subsystem
(designated by the shaded regions in each panel). 
  \label{fig:sllscldy}}
\end{figure}

\begin{deluxetable}{lccc}
\tablewidth{0pc}
\tablecaption{ELEMENTAL ABUNDANCES FOR SLLS AT $z \approx 3.19$\label{tab:xh319}}
\tabletypesize{\footnotesize}
\tablehead{\colhead{Ion} &
\colhead{[X/H]} & \colhead{[X/Si$^+$]}}
\startdata
\cutinhead{SLLS$_a$}
C$^{+}$ & $-2.47\pm0.10$ & $+ 0.16\pm0.10$ \\
O$^{0}$ & $-2.26\pm0.10$ & $+ 0.37\pm0.10$ \\
Al$^{+}$ & $-3.04\pm0.12$ & $-0.40\pm0.10$ \\
Si$^{+}$ & $-2.64\pm0.10$ & $$ \\
Fe$^{+}$ & $<-2.43$ & $<+ 0.20$ \\
\cutinhead{SLLS$_b$}
C$^{+}$ & $>-1.64$ & $>-0.22$ \\
O$^{0}$ & $>-2.20$ & $>-0.79$ \\
Al$^{+}$ & $-1.72\pm0.10$ & $-0.30\pm0.10$ \\
Si$^{+}$ & $-1.42\pm0.10$ & $$ \\
Fe$^{+}$ & $-1.61\pm0.10$ & $-0.19\pm0.10$ \\
\cutinhead{SLLS$_c$}
C$^{+}$ & $>-1.21$ & $>+ 0.02$ \\
Al$^{+}$ & $-1.58\pm0.11$ & $-0.35\pm0.10$ \\
Si$^{+}$ & $-1.24\pm0.10$ & $$ \\
Fe$^{+}$ & $-1.42\pm0.10$ & $-0.19\pm0.10$ \\
\enddata
\tablenotetext{a}{Assumes a Cloudy photoionization model with $\mnhi = 10^{19.8} \cm{-2}$, [M/H] = $-2.00$ and log $U = -2.8\pm 0.2$\,dex}
\tablenotetext{b}{Assumes a Cloudy photoionization model with $\mnhi = 10^{19.8} \cm{-2}$, [M/H] = $-1.50$ and log $U = -4.0\pm 0.2$\,dex}
\tablenotetext{c}{Assumes a Cloudy photoionization model with $\mnhi = 10^{19.0} \cm{-2}$, [M/H] = $-0.50$ and log $U = -2.9\pm 0.2$\,dex}
\tablecomments{In all cases, we have assumed a minimum error of 0.1\,dex
due to systematic errors in the photoionization modeling.}
\end{deluxetable}

\begin{deluxetable}{cccc}
\tablewidth{0pc}
\tablecaption{SUMMARY OF PROPERTIES FOR THE SLLS AT $z \approx 3.19$\label{tab:sllssumm}}
\tabletypesize{\footnotesize}
\tablehead{\colhead{Property} &\colhead{a} & \colhead{b} & \colhead{c} }
\startdata
log (\nhi/$\cm{-2}$) &$19.75\pm0.15$&$19.80\pm0.15$&$19.10\pm0.15$\\
log U &$-2.8\pm 0.2$&$-4.0\pm 0.2$&$-2.9\pm 0.2$\\
$\log (1-x)^a$ &$-0.66_{-0.17}^{+0.17}$&$-0.09_{-0.02}^{+0.05}$&$-1.22_{-0.19}^{+0.20}$\\
$\log (N_{\rm H}/\cm{-2})$ &$20.41\pm0.23$&$19.89\pm0.15$&$20.32\pm0.25$\\
$\lbrack$Si/H]&$ -2.64 \pm 0.16$&$ -1.42 \pm 0.15$&$ -1.20 \pm 0.17$\\
$\lbrack$O/H]&$ -2.26 \pm 0.25$&$> -2.20$&$$\\
$\lbrack$C/H]&$ -2.47 \pm 0.15$&$> -1.64$&$> -1.24$\\
$\lbrack$Fe/H]&$$&$ -1.61 \pm 0.15$&$ -1.40 \pm 0.17$\\
\enddata
\tablenotetext{a}{The ionization fraction $x$ is defined as H$^+$/H.}
\tablecomments{Abundances assume the photoionization models as
described in Table~\ref{tab:xh319}.}
 
\end{deluxetable}

\subsection{Photoionization Modeling}

We have performed a similar analysis for the SLLSs at 
$z \approx 3.19$ toward \qso, adopting Cloudy models
with larger \nhi\ values and a EUVB radiation field at the
appropriate redshift.
Figure~\ref{fig:sllscldy} summarizes the analysis.  Note that
for \aslls\ we have adopted upper limits to the $\rm C^{+3}$
and $\rm Si^{+3}$ column densities by integrating the apparent
optical depth over the same velocity intervals as the observed
low-ion absorption.  There is additional high-ion absorption at
$\delta v \approx -1225 \mkms$ relative to $z=3.190$ (Figure~\ref{fig:sllsmtl})
which must be arising in a more highly ionized cloud.  We assume this
gas does not contribute significantly to the observed \ion{H}{1} absorption.
In contrast to these high-ions, we do
detect strong \ion{Si}{3}~1206 absorption which traces the line-profiles
of the low-ion transitions.
Attributing this gas to the same phase as the low-ions, we find
$\log U_a = -2.8 \pm 0.2$\,dex.  
If we have overestimated the ionization parameter by making this
association, then we will have underestimated
the [Si/H] abundance by $\approx 0.3$\,dex.  The 
[O/H] estimate (based on O$^0$/H$^0$), however, establishes that
\aslls\ has a metallicity below $1/100$ solar.

The absence of significant high-ion 
absorption at any velocity near \bslls\ implies $\log U_b < -4$. To
be conservative,  
we adopt this limit as the central value in the following analysis
noting that lower values give very similar results.
Finally, we estimate $\log U_c = -2.9 \pm 0.2$\,dex based on the observed
ionic ratios of C, Si, and Al.
The absolute and relative abundances of the SLLSs are
reported in Table~\ref{tab:xh319}.

%%%%%%%%%%%%%%%%%%%%%%%%%%%%%%%%%%%%%%%%%%%%%%%%%%%%%%%%%%%%%%%%%%%%%%%%
%\bibliographystyle{/u/xavier/paper/Bibli/apj}
%\bibliography{/u/xavier/paper/Bibli/allrefs}

%\input{../Tables/tab_xh355.tex}

%\input{../Tables/tab_llssumm.tex}

%\input{../Tables/tab_z319clm.tex}

\clearpage

\end{document}